\documentclass[twocolumn,twocolappendix]{aastex631}

\usepackage{amsmath}	

\shorttitle{High Ionization Emission Lines in $z>4$ Galaxies}
\shortauthors{Tang et al.}

\begin{document}

\title{JWST/NIRSpec Observations of High Ionization Emission Lines in Galaxies at High Redshift}

\author[0000-0001-5940-338X]{Mengtao Tang}
\affiliation{Steward Observatory, University of Arizona, 933 N Cherry Ave, Tucson, AZ 85721, USA}
\email{tangmtasua@arizona.edu}

\author{Daniel P. Stark}
\affiliation{Department of Astronomy, University of California, Berkeley, Berkeley, CA 94720, USA}

\author{Ad\`ele Plat}
\affiliation{Institute for Physics, Laboratory for Galaxy Evolution and Spectral Modelling, Ecole Polytechnique Federale de Lausanne, Observatoire de Sauverny, Chemin Pegasi 51, CH-1290 Versoix, Switzerland}

\author{Anna Feltre}
\affiliation{INAF - Osservatorio Astrofisico di Arcetri, Largo E. Fermi 5, I-50125, Firenze, Italy}

\author{Harley Katz}
\affiliation{Department of Astronomy \& Astrophysics, University of Chicago, 5640 S Ellis Avenue, Chicago, IL 60637, USA}

\author{Peter Senchyna}
\affiliation{The Observatories of the Carnegie Institution for Science, 813 Santa Barbara Street, Pasadena, CA 91101, USA}

\author{Charlotte A. Mason}
\affiliation{Cosmic Dawn Center (DAWN)}
\affiliation{Niels Bohr Institute, University of Copenhagen, Jagtvej 128, 2200 Copenhagen N, Denmark}

\author{Lily Whitler}
\affiliation{Steward Observatory, University of Arizona, 933 N Cherry Ave, Tucson, AZ 85721, USA}

\author{Zuyi Chen}
\affiliation{Steward Observatory, University of Arizona, 933 N Cherry Ave, Tucson, AZ 85721, USA}

\author{Michael W. Topping}
\affiliation{Steward Observatory, University of Arizona, 933 N Cherry Ave, Tucson, AZ 85721, USA}



\begin{abstract}
{\it JWST} spectroscopy has built large emission line samples at $z\gtrsim 4$, but it has yet to confidently reveal many galaxies with the hard radiation fields commonly associated with AGN photoionization.  
While this may indicate a weaker UV ionizing spectrum in many $z>4$ AGNs or obscuration from dense neutral gas and dust, the complete picture remains unclear owing to the small number of deep rest-UV spectra.
Here we characterize the strength of high ionization lines in $53$ new galaxies observed with NIRSpec $R=2700$ grating spectroscopy.
We present new detections of narrow N~{\small V}~$\lambda1240$ in two galaxies. 
One is a previously-confirmed $z=6.98$ Little Red Dot (LRD) with broad H$\beta$, and the other is a $z=8.72$ galaxy with a narrow line spectrum. 
Neither source exhibits C~{\small IV} or He~{\small II} emission, indicating large N~{\small V}/C~{\small IV} and N~{\small V}/He~{\small II} ratios that may reflect a combination of nitrogen-enhancement and resonant scattering effects. 
We investigate the incidence of narrow high ionization lines in a large database of $851$ NIRSpec grating spectra, and we separately quantify the fraction of LRDs with narrow high ionization UV emission lines. 
Our results likely suggest that hard radiation fields are indeed present in a small subset of LRDs ($12.5^{+23.7}_{-10.4}\%$) and UV-selected galaxies ($2.2^{+1.7}_{-1.0}\%$) at $z>4$. 
The identification of narrow high ionization lines in the population of LRDs with strong Balmer absorption suggests the dense neutral hydrogen gas may not uniformly cover the nucleus. 
The strong N~{\small V} (coupled with weak C~{\small IV} and He~{\small II}) suggests that efforts to identify high ionization lines should extend down in wavelength to the N~{\small V} doublet.
\end{abstract}




\section{Introduction} \label{sec:introduction}

One decade ago, the first rest-frame ultraviolet (UV) metal emission lines (C~{\small III}], C~{\small IV}) were detected in bright $z\simeq6-7$ galaxies \citep{Stark2015a,Stark2015b,Stark2017,Laporte2017,Mainali2017,Hutchison2019,Topping2021}. 
The equivalent widths (EWs) were found to be over an order of magnitude greater than what is common at $z\simeq 2-3$, suggesting that there might be significant evolution in the gas conditions and radiation field at $z\gtrsim 6$. 
The origin of the hard radiation field was not clear with ground-based data, with some arguing that an active galactic nucleus (AGN) was responsible \citep{Nakajima2018} and others suggesting the presence of low metallicity massive stars \citep[e.g.,][]{Mainali2017,Stark2017}. 
Regardless of the powering mechanism, the discovery of such strong emission (EW $>20-40$~\AA) led to the suggestion that the UV metal lines would be readily detectable in galaxies at $z\simeq10-15$ with NIRSpec \citep{Jakobsen2022,Boker2023} spectroscopy, providing a viable path toward characterizing the stars, gas and ionizing sources in the most distant galaxies that {\it JWST} \citep{Gardner2023,Rigby2023} discovers.  

This prediction has been borne out with {\it JWST} rest-frame UV spectroscopy. 
In the last year, we have seen the redshift frontier extended to $z\simeq14$ \citep{Carniani2024}, with UV metal lines often providing our best path toward characterization of the most distant galaxies that have been discovered \citep{Bunker2023,Carniani2024,Castellano2024,DEugenio2024,Hsiao2024,Napolitano2025}. 
Deeper spectroscopy has been obtained of several of the brightest $z\gtrsim6$ galaxies showing intense C~{\small III}] and C~{\small IV} emission \citep{Topping2024,Topping2025a}. 
These spectra show a suite of intense lines from species with ionization energies up to $54$~eV (the He$^+$ edge), but yet higher ionization lines are not seen (i.e. N~{\small V}, [Ne~{\small IV}], [Ne~{\small V}]). 
The implied radiation field appears consistent with stellar photoionization. 
While it is plausible that AGN contribute significantly to these lines, there is no clear indication from the observed extreme UV (EUV) line ratios. 

If the majority of galaxies with strong UV metal lines at $z\gtrsim6$ are (mostly) dominated by stars, the question remains whether a separate population of AGNs with high ionization line\footnote{Throughout the paper we refer to high ionization lines as lines from species with ionization energies above the He$^{+}$ ionizing edge ($54$~eV), e.g., N~{\scriptsize V}, [Ne~{\scriptsize IV}], and [Ne~{\scriptsize V}].} emission is also present. 
While large samples of $z\gtrsim4$ broad-line (Type I) AGNs (BL AGNs) have been revealed with {\it JWST} spectroscopy \citep[e.g.,][]{Kocevski2023,Kokorev2023,Ubler2023,Furtak2024,Greene2024,Maiolino2024b,Matthee2024,Ubler2024,Kocevski2025}, there have been relatively few robust detections of very high ionization lines \citep[e.g.,][]{Brinchmann2023,Chisholm2024,Mazzolari2024b,Silcock2024,Scholtz2025,Treiber2025} commonly seen in narrow line (Type II) AGN (NL AGN) spectra \citep[e.g.,][]{Zakamska2003,Gilli2010,Hainline2011,Peca2025,Trakhtenbrot2025}. 
Additionally the rest-frame UV spectra of the BL AGN discovered with {\it JWST} have thus far not revealed broad permitted high ionization line emission in the rest-frame UV. 
It has been suggested that the absence of high ionization lines in the Type I AGN may be a consequence of super-Eddington accretion, with photon trapping in the thick accretion disk leading to an intrinsically weaker UV spectrum \citep[e.g.,][]{Pognan2020,Lambrides2024}. 
Additionally, if the BL AGN are surrounded by extremely dense neutral gas with a near-unity covering fraction and the $n=2$ level of hydrogen populated \citep[e.g.,][]{deGraaff2025a,Inayoshi2025,Ji2025,Naidu2025}, the emergent UV lines would face significant opacity, and the ionizing continuum would be attenuated before reaching the narrow line emitting region. 
Both factors may contribute to the absence of high ionization line emission.

Progress in characterizing the UV radiation field of $z\gtrsim4$ AGNs will only be possible with larger samples of deep rest-frame UV spectra. 
Toward this end, we present a search for high ionization line emission in $851$ $z\gtrsim4$ galaxies with NIRSpec grating spectra ($R=1000$ or $R=2700$), including a sample of $18$ $z\gtrsim4$ galaxies confirmed as BL AGN with {\it JWST} spectroscopy. 
This database includes new observations from a NIRSpec program (GO 4287, PI: Mason) using the high resolution ($R=2700$) G140H/F100LP grating to sample the rest-frame UV of $z\gtrsim6$ galaxies in the Extended Groth Strip (EGS) field \citep{Davis2007}. 
These new observations have revealed two detections of narrow emission lines near the N~{\small V} resonance. 
One is found in a BL AGN at $z=6.98$ (CEERS-7902, \citealt{Labbe2023,Kocevski2025}), whereas the spectral energy distribution (SED) of the other is suggestive of a fairly typical star forming galaxy at $z=8.72$ (CEERS-1025, \citealt{Nakajima2023,Tang2023}). 
In this paper, we use our full database to assess the fraction of $z\gtrsim4$ galaxies showing high ionization lines, as might be expected from an AGN power law spectrum with significant UV continuum output. 
We also quantify the strength of high ionization line emission (narrow and broad) in $z\gtrsim4$ Little Red Dots (LRDs; \citealt{Matthee2024}) that have been confirmed to show broad line hydrogen lines in the rest-frame optical. 
We present and describe new rest-frame UV spectra for two LRDs (CEERS-7902 and CEERS-10444).   

The organization of this paper is as follows. 
In Section~\ref{sec:data}, we describe the GO 4287 observations and the data analysis. 
We then characterize the spectroscopic properties of the two likely N~{\small V} emitters (CEERS-1025, CEERS-7902) and a LRD (CEERS-10444) from GO 4287 in Section~\ref{sec:objects}. 
In Section~\ref{sec:line_ratios}, we compare the implied N~{\small V} line ratios to expectations from photoionization models. 
We then search for high ionization line emission in the full archival JWST/NIRSpec sample at $z>4$ in Section~\ref{sec:archive} and discuss implications of the census in Section~\ref{sec:discussion}.
Finally, we summarize our conclusions in Section~\ref{sec:summary}. 
Throughout the paper, we adopt a $\Lambda$-dominated, flat universe with $\Omega_{\Lambda}=0.7$, $\Omega_{\rm{M}}=0.3$, and $H_0=70$~km~s$^{-1}$~Mpc$^{-1}$. 
All magnitudes are quoted in the AB system \citep{Oke1983} and all EWs are quoted in the rest frame.


\section{Spectroscopic Data and Analysis} \label{sec:data}


\begin{figure*}
\includegraphics[width=\linewidth]{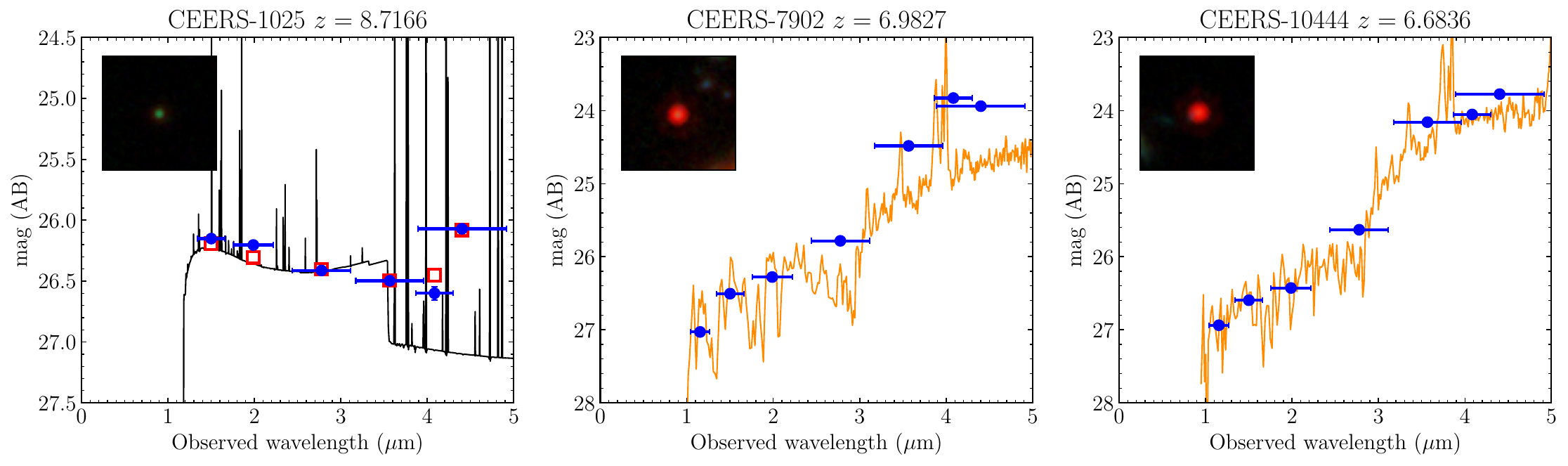}
\caption{SEDs of the two N~{\scriptsize V} emitters CEERS-1025 (left) and CEERS-7902 (LRD; middle), as well as CEERS-10444 (LRD; right). Observed photometry is shown by blue circles. Left: We show the \texttt{BEAGLE} model \citep{Chevallard2016} spectrum of CEERS-1025 as the black line and the synthetic photometry as red squares. Middle and right: We show the NIRSpec prism spectra of CEERS-7902 and CEERS-10444 obtained from RUBIES \citep{deGraaff2025b} as the orange lines. We insert the postage stamp ($1.5''\times1.5''$) of each object. The RGB image are composed of images of F115W, F200W, and F444W bands.}
\label{fig:SED}
\end{figure*}

In this section, we describe the JWST/NIRSpec spectroscopic data obtained from the Cycle 2 program GO 4287 (PI: C. Mason). 
This program targets high redshift galaxies in the NIRCam \citep{Rieke2005,Rieke2023} imaging footprint of the EGS field.   
The spectra were obtained using JWST/NIRSpec in multi-object spectroscopy mode in 2024 March. 
We observed three micro-shutter assembly (MSA; \citealt{Ferruit2022}) mask configurations. 
The primary targets placed on these configurations are galaxies in candidate overdensities that may trace ionized bubbles. 
We provide a brief summary of the observations here. 
A full description and catalog will be presented in Whitler et al. (2025, in prep).

The observations were conducted with MSA configurations using the high-resolution ($R=2700$) grating/filter pair G140H/F100LP (targeting the rest-frame UV) and the medium-resolution ($R=1000$) grating/filter pair G395M/F290LP (targeting the rest-frame optical).
The total exposure time was $14005$~s for G140H/F100LP and $3501$~s for G395M/F290LP. 
For each slit on the MSA, we used a $3$-shutter slitlet and observed with $3$-nod pattern for dithering. 

The 2D NIRSpec spectra were reduced following the procedures described in \citet{Topping2025a} using the standard JWST data reduction pipeline\footnote{\url{https://github.com/spacetelescope/jwst}} \citep{Bushouse2024}. 
The 1D spectra are then extracted from the reduced 2D spectra using a boxcar extraction. 
The typical extraction aperture is $5$~pixels ($\sim0.5$~arcsec in the spatial direction). 
For the G140H/F100LP spectra, we find that one resolution element corresponds to $\simeq111$~km~s$^{-1}$. 
In these spectra, the median $3\sigma$ limiting line flux is $3.5\times10^{-19}$~erg~s$^{-1}$~cm$^{-2}$ for an unresolved emission line (in the spectral direction) produced by a point source. 
At this limiting flux, we can detect weak emission lines (EW $\simeq5$~\AA) in the rest-frame UV for moderately faint continuum sources ($H=27.0$) at $z\simeq8$. 
For G395M/F290LP spectra, the spectral resolution is closer to $\simeq300$~km~s$^{-1}$ and the median $3\sigma$ limiting line flux is $4.6\times10^{-19}$~erg~s$^{-1}$~cm$^{-2}$. 

For this analysis, we have also updated our reductions of spectra of other targets in the EGS field, building on our earlier work \citep{Tang2023,Chen2024,Tang2024c}. 
In particular, we add in new data from the Red Unknowns: Bright Infrared Extragalactic Survey (RUBIES, GO 4233, PIs: A. de Graaff \& G. Brammer; \citealt{deGraaff2025b}) program, reduced following the same procedures we used to process the GO 4287 data.
The RUBIES observations target the EGS and the Ultra Deep Survey (UDS) fields with {\it JWST}/NIRSpec, using both the low resolution ($R\sim100$) prism (covering $0.6-5.3\ \mu$m) and medium resolution grating/filter pair G395M/F290LP (covering $2.8-5.2\ \mu$m). 
We refer readers to \citet{deGraaff2025b} for a detailed description. 

In this paper, we present the spectra of two galaxies (CEERS-1025 and CEERS-7902) for which we have detected an emission line near the N~{\small V} doublet.
CEERS-1025 was first observed by the Cosmic Evolution Early Release Science\footnote{\url{https://ceers.github.io/}} (CEERS, ERS 1345, PI: S. Finkelstein; \citealt{Finkelstein2025}) program, providing $R=1000$ spectra targeting from the rest-frame UV to optical. 
To maximize the S/N, we generate a composite G395M spectrum for CEERS-1025 by stacking the GO 4287 and CEERS $R=1000$ grating spectrum \citep{Tang2023}, with the weighting set by the exposure times. 
CEERS-7902 was also observed by RUBIES (and discussed in \citealt{Kocevski2025,Wang2025}).
We generate a composite G395M spectrum for CEERS-7902 by stacking the GO 4287 and RUBIES grating spectra. 
We also present new spectra of a third galaxy, CEERS-10444, a LRD that was also observed in the RUBIES program. 
We describe our new rest-frame UV spectrum and stack our rest-frame optical G395M spectrum with that obtained in the RUBIES program. 
In the following analysis, we will use GO 4287 G140H spectra and the composite G395M spectra.

We perform the emission line measurements for CEERS-1025, CEERS-7902, and CEERS-10444 as follows. 
For emission lines detected with S/N $>5$, we measure the line fluxes, centroids, widths, and EWs of each galaxy by fitting the line profiles and nearby continua with Gaussian functions on top of linear functions. 
For isolated emission lines, we use a single Gaussian. 
For emission lines that are close in wavelength (i.e., H$\gamma$ and [O~{\small III}]~$\lambda4363$), we fit with multiple Gaussians simultaneously. 
In the case of emission lines with broad components, we fit the line with a narrow Gaussian on top of a broad Gaussian. 
For emission lines detected with lower S/N ($<5$), we calculate the line fluxes using direct integration. 
We note that Gaussian fit and direct integration result in similar line fluxes for emission lines with S/N $>5$.
We evaluate the uncertainties of the line fluxes and EWs using the following procedure. 
We resample the flux densities of each spectrum $1000$ times by taking the observed flux densities as the mean values and the errors as standard deviations. 
Then we compute the line fluxes and EWs from the resampled spectra of each galaxy using the same approach described above. 
We take the standard deviation of these measurements as the uncertainty. 
The same analysis is presented to the other $55$ $z>4$ galaxies in the GO 4287 spectra. 
While these will be presented in more detail in a later work (Whitler et al. 2025, in prep), we will compare the line ratios of the N~{\small V} emitters to those in the full sample (in addition to other sources in the archive) to place the likely N~{\small V} detections in context.


\section{High Ionization Emission in GO 4287} \label{sec:objects}

We present new NIRSpec observations of CEERS-1025 (Section~\ref{sec:CEERS-1025}), CEERS-7902 (Section~\ref{sec:CEERS-7902}), 
and CEERS-10444 (Section~\ref{sec:CEERS-10444}) from GO 4287.
We first discuss what was known about both galaxies from earlier {\it JWST} observations before describing the new emission line detections.
Then we discuss the interpretation of line ratios between N~{\small V} and other UV lines in Section~\ref{sec:line_ratios}.


\begin{deluxetable}{ccc}
\tablecaption{Systemic redshifts and rest-frame UV to optical emission line fluxes ($\times10^{-19}$~erg~s$^{-1}$~cm$^{-2}$) of the two N~{\scriptsize V} emitters. We show $3\sigma$ upper limits for non-detections.}
\tablehead{
 & CEERS-1025 & CEERS-7902
}
\startdata
$z_{\rm sys}$ & $8.7166$ & $6.9827$ \\
\hline
N~{\scriptsize V}~$\lambda1239$ & $<5.2$ & $12.0\pm1.5$ \\
N~{\scriptsize V}~$\lambda1243$ & $14.0\pm2.3$ & $<3.2$ \\
{[}N~{\scriptsize IV}]~$\lambda1483$ & $<4.7$ & $<2.5$ \\
N~{\scriptsize IV}]~$\lambda1486$ & $<5.0$ & $<2.4$ \\
C~{\scriptsize IV}~$\lambda1548$ & $<5.3$ & $<2.2$ \\
C~{\scriptsize IV}~$\lambda1551$ & $<4.7$ & $<2.4$ \\
He~{\scriptsize II}~$\lambda1640$ & $<5.3$ & $<2.5$ \\
O~{\scriptsize III}]~$\lambda1661$ & $<5.6$ & $<2.2$ \\
O~{\scriptsize III}]~$\lambda1666$ & $<5.0$ & $<2.4$ \\
N~{\scriptsize III}]~$\lambda1746$ & $<5.9$ & - \\
N~{\scriptsize III}]~$\lambda1748$ & $<5.6$ & - \\
{[}C~{\scriptsize III}]~$\lambda1907$ & - & $<2.9$ \\
C~{\scriptsize III}]~$\lambda1909$ & - & $3.5\pm1.2$ \\
{[}O~{\scriptsize II}]~$\lambda3728$ & $8.9\pm2.1$ & $2.8\pm1.5$ \\
{[}Ne~{\scriptsize III}]~$\lambda3869$ & $3.0\pm1.2$ & $5.5\pm1.3$ \\
H$\gamma$ (total) & $6.9\pm1.9$ & $11.2\pm3.0$ \\
H$\gamma$ (narrow) & - & $4.0\pm1.3^{\rm a}$ \\
H$\gamma$ (broad) & - & $7.2\pm2.7^{\rm b}$ \\
{[}O~{\scriptsize III}]~$\lambda4363$ & $3.6\pm1.4$ & $7.3\pm1.4$ \\
H$\beta$ (total) & $14.5\pm2.4$ & $50.2\pm3.9$ \\
H$\beta$ (narrow) & - & $5.9\pm1.1^{\rm a}$ \\
H$\beta$ (broad) & - & $44.4\pm3.7^{\rm b}$ \\
{[}O~{\scriptsize III}]~$\lambda4959$ & $21.9\pm2.3$ & $19.0\pm1.3$ \\
{[}O~{\scriptsize III}]~$\lambda5007$ & $63.5\pm2.8$ & $57.7\pm1.7$ \\
He~{\scriptsize I}~$\lambda5877$ & - & $5.7\pm2.1$ \\
{[}O~{\scriptsize I}]~$\lambda6302$ & - & $4.4\pm1.1$ \\
\enddata
\tablecomments{a: Narrow component of the H$\gamma$ or H$\beta$ emission line of CEERS-7902. b: Broad component of the H$\gamma$ or H$\beta$ of CEERS-7902.}
\label{tab:line_flux}
\end{deluxetable}


\begin{deluxetable}{ccc}
\tablecaption{Rest-frame UV to optical emission line EWs (\AA) and $3\sigma$ upper limits of the two N~{\scriptsize V} emitters.}
\tablehead{
 & CEERS-1025 & CEERS-7902
}
\startdata
N~{\scriptsize V}~$\lambda1239$ & $<2.7$ & $27.9\pm3.4$ \\
N~{\scriptsize V}~$\lambda1243$ & $7.0\pm1.1$ & $<7.7$ \\
{[}N~{\scriptsize IV}]~$\lambda1483$ & $<3.1$ & $<8.6$ \\
N~{\scriptsize IV}]~$\lambda1486$ & $<3.3$ & $<8.6$ \\
C~{\scriptsize IV}~$\lambda1548$ & $<3.9$ & $<8.7$ \\
C~{\scriptsize IV}~$\lambda1551$ & $<3.3$ & $<9.2$ \\
He~{\scriptsize II}~$\lambda1640$ & $<4.8$ & $<14$ \\
O~{\scriptsize III}]~$\lambda1661$ & $<5.3$ & $<11$ \\
O~{\scriptsize III}]~$\lambda1666$ & $<4.7$ & $<13$ \\
N~{\scriptsize III}]~$\lambda1746$ & $<5.2$ & - \\
N~{\scriptsize III}]~$\lambda1748$ & $<5.1$ & - \\
{[}C~{\scriptsize III}]~$\lambda1907$ & - & $<19$ \\
C~{\scriptsize III}]~$\lambda1909$ & - & $22.5\pm7.7$ \\
{[}O~{\scriptsize II}]~$\lambda3728$ & $39\pm10$ & $10\pm6$ \\
{[}Ne~{\scriptsize III}]~$\lambda3869$ & $16\pm6$ & $20\pm5$ \\
H$\gamma$ (total) & $75\pm21$ & $33\pm9$ \\
H$\gamma$ (narrow) & - & $12\pm4^{\rm a}$ \\
H$\gamma$ (broad) & - & $21\pm8^{\rm b}$ \\
{[}O~{\scriptsize III}]~$\lambda4363$ & $39\pm15$ & $22\pm4$ \\
H$\beta$ (total) & $200\pm34$ & $141\pm11$ \\
H$\beta$ (narrow) & - & $17\pm3^{\rm a}$ \\
H$\beta$ (broad) & - & $124\pm10^{\rm b}$ \\
{[}O~{\scriptsize III}]~$\lambda4959$ & $331\pm35$ & $52\pm4$ \\
{[}O~{\scriptsize III}]~$\lambda5007$ & $1004\pm44$ & $155\pm5$ \\
He~{\scriptsize I}~$\lambda5877$ & - & $17\pm6$ \\
{[}O~{\scriptsize I}]~$\lambda6302$ & - & $11\pm3$ \\
\enddata
\tablecomments{a: Narrow component of the H$\gamma$ or H$\beta$ emission line of CEERS-7902. b: Broad component of the H$\gamma$ or H$\beta$ of CEERS-7902.}
\label{tab:line_ew}
\end{deluxetable}


\begin{figure*}
\includegraphics[width=\linewidth]{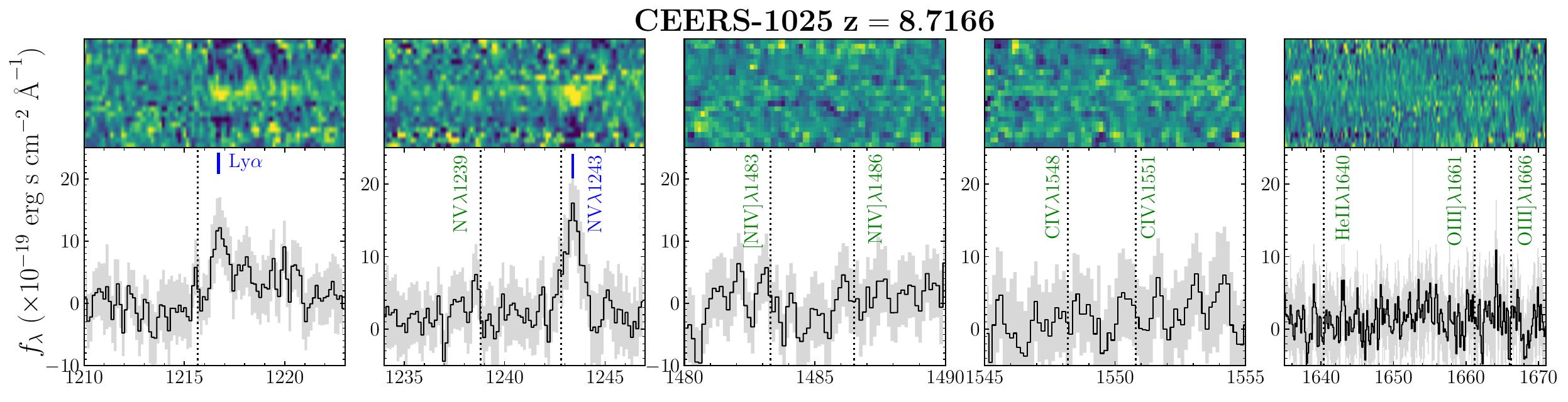}
\includegraphics[width=\linewidth]{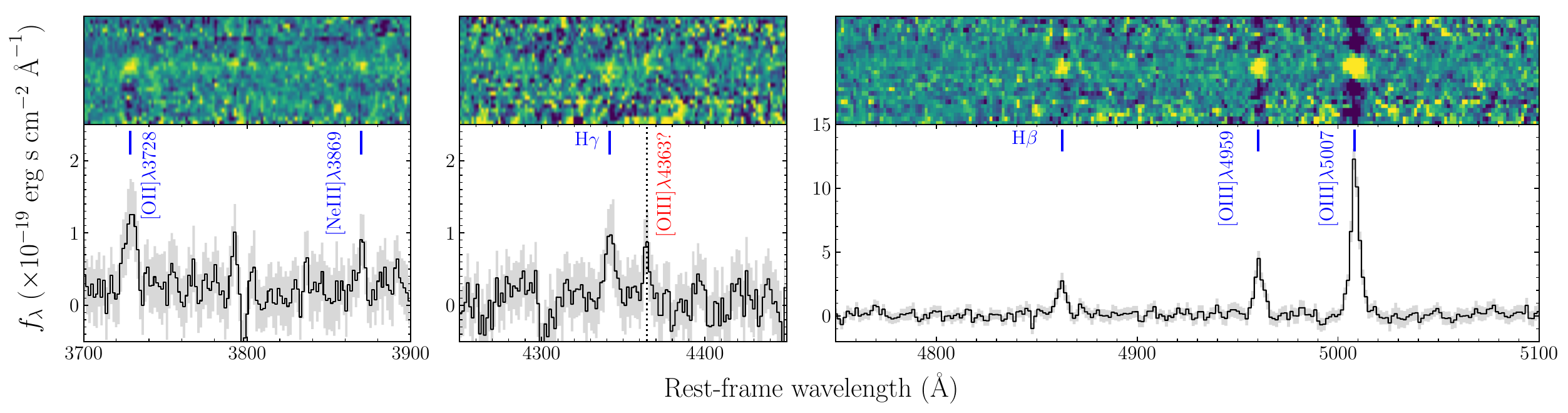}
\caption{JWST/NIRSpec grating spectra of the N~{\scriptsize V} emitter CEERS-1025. The top panels show the GO 4287 G140H spectra and the bottom panels show the composite GO 4287 and CEERS G395M spectra. We overplot the expected positions of emission lines from the systemic redshift ($z_{\rm sys}=8.7166$) as black dotted lines. Detected emission lines are marked by blue solid lines. [O~{\scriptsize III}]~$\lambda4363$ is tentatively (S/N $=2.7$) detected and marked by red text. Non-detections (N~{\scriptsize V}~$\lambda1239$, N~{\scriptsize IV}], C~{\scriptsize IV}, He~{\scriptsize II}, O~{\scriptsize III}]) are shown by green text.}
\label{fig:CEERS-1025_spec}
\end{figure*}

\subsection{Spectroscopy of $z=8.72$ galaxy CEERS-1025} \label{sec:CEERS-1025}

CEERS-1025 (R.A. $=214.967594$, Decl. $=+52.933005$) is a bright galaxy ($m_{\rm F150W}=26.2$, M$_{\rm UV}=-21.2$) at $z=8.72$. 
The redshift of this object was first confirmed from the CEERS observations using the medium-resolution NIRSpec grating (G140M/F100LP, G235M/F170LP, G395M/F290LP) \citep[e.g.,][]{Nakajima2023,Tang2023}. 
These observations cover rest-frame wavelengths of $1010-1120$~\AA\ and $1245-5350$~\AA, with a chip gap leading to the missing spectral coverage between Ly$\alpha$ and N~{\small V} emission. 

As described in \citet{Tang2023}, the NIRCam SED from CEERS (left panel of Figure~\ref{fig:SED}) reveals a blue UV slope  ($\beta=-2.5$), similar to typical galaxies at $z\simeq8$. 
There is a clear flux density upturn in the F444W filter (relative to the neighboring shorter wavelength filters), suggesting either large EW [O~{\small III}] emission (as expected for galaxies dominated by young stellar populations) or a reddened rest-frame optical continuum (as seen in many broad line AGN seen with {\it JWST}). 
The CEERS G395M spectrum confirms the former case, revealing very strong line emission ([O~{\small III}]+H$\beta$ EW $=1493\pm93$~\AA) seen in many $z\gtrsim7$ galaxies.  

The new G140H spectrum presented in this paper covers rest-frame wavelengths between $1000$~\AA\ and $1880$~\AA\ for CEERS-1025 (Figure~\ref{fig:CEERS-1025_spec}), providing deeper and higher resolution rest-frame UV coverage and the first useful constraints blueward of $1245$~\AA. 
We present the new rest-frame UV spectrum in Section~\ref{sec:CEERS-1025_uv} before briefly discussing the newly-acquired rest-frame optical spectrum in Section~\ref{sec:CEERS-1025_opt}.
We will adopt a systemic redshift $z_{\rm sys}=8.7166$ derived from the [O~{\small III}] doublet and H$\beta$ emission lines in the GO 4287 G395M spectrum, which is consistent with the redshift reported in \citet{Nakajima2023} and \citet{Tang2023}.

\subsubsection{Rest-Frame UV Spectrum of CEERS-1025} \label{sec:CEERS-1025_uv}

We visually search the rest-frame UV spectrum of CEERS-1025 (top panel of Figure~\ref{fig:CEERS-1025_spec}) for emission lines. 
One feature is detected at an observed frame wavelength of $11822.9$~\AA\ (S/N $=5$), near the rest-frame wavelength of Ly$\alpha$. 
The only other robust emission feature detected in the rest-frame UV spectrum is at observed wavelength of $12081.6$~\AA\ (S/N $=6$), close to the expected wavelengths of the N~{\small V}~$\lambda\lambda1239,1243$ resonant doublet. 
If this feature is nebular N~{\small V} emission, it would suggest CEERS-1025 powers a hard radiation field, with a supply of photons with energies $>77$~eV capable of producing N~{\small V} ions.

The peak flux of the Ly$\alpha$ emission line is redshifted by $+272\pm33$~km~s$^{-1}$ from Ly$\alpha$ resonance. 
The core of the Ly$\alpha$ line is narrow, but there is a red tail of emission extending to $\sim 1000$~km~s$^{-1}$. 
The redshifted peak velocity and asymmetric profile of Ly$\alpha$ is consistent with the presence of outflowing neutral gas in CEERS-1025, as is commonly seen in high redshift galaxies. 
We cannot rule out the possibility that the extended red wing is associated with the broad line region of an AGN (with the blue side scattered by the IGM), but since no similar broad profile is seen for H$\beta$, we will assume that the asymmetric Ly$\alpha$ profile is more likely driven by backscattering in a neutral outflow. 
The Ly$\alpha$ emission line of CEERS-1025 (and implications for the IGM) will be discussed in detail in Whitler et al. (2025, in prep). 

The N~{\small V} line is a doublet ($\lambda=1238.8$~\AA\ and $\lambda=1242.8$~\AA), but only one component is detected in our spectrum. 
The detected feature is narrow, with no clear broad component present in the existing spectrum. 
After de-convolving the instrument resolution (assuming a point source), the detected line has a full width at half maximum (FWHM) of $208\pm28$~km~s$^{-1}$. 
The emission line is closest to the red component of the doublet, with a peak flux redshifted by $+144\pm32$~km~s$^{-1}$ from line center. 
In what follows, we will assume the feature is N~{\small V}~$\lambda1243$, but we note that we cannot immediately rule out that we are seeing the blue component at a very high velocity ($+1113\pm32$~km~s$^{-1}$) from line center. 
As with Ly$\alpha$, the redshifted N~{\small V} profile is consistent with backscattering off of outflowing gas, but in this case, the presence of N~{\small V} resonant scattering would indicate that the outflowing material has a very highly ionized component capable of scattering N~{\small V} photons. 
We measure a line flux of $1.40\pm0.23\times10^{-18}$~erg~s$^{-1}$~cm$^{-2}$ and an associated rest-frame EW of $7.0\pm1.1$~\AA. 
A deeper spectrum detecting both components of the N~{\small V} doublet will provide a better interpretation of the nature of N~{\small V}.


\begin{figure}
\includegraphics[width=\linewidth]{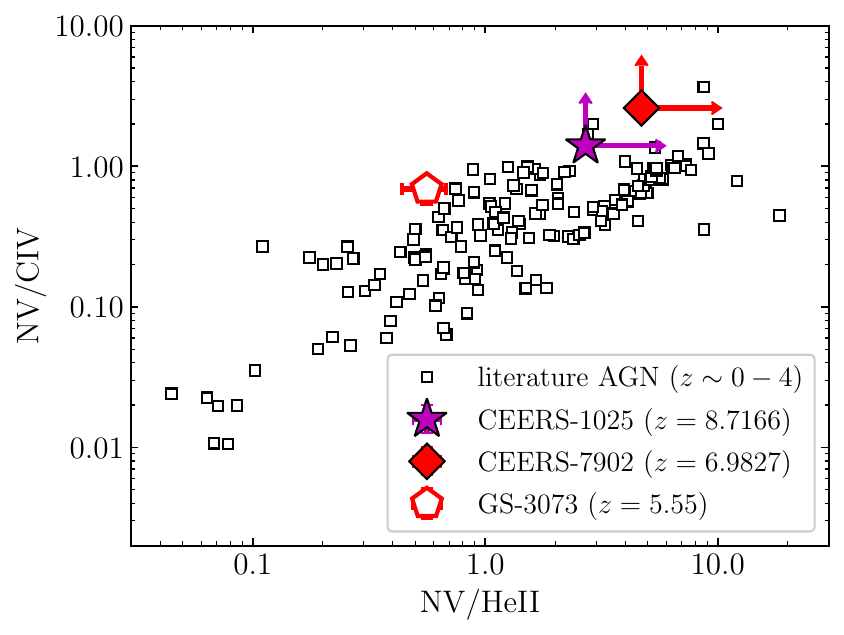}
\caption{N~{\scriptsize V}~$\lambda\lambda1239,1243$/C~{\scriptsize IV}~$\lambda\lambda1548,1551$ versus N~{\scriptsize V}~$\lambda\lambda1239,1243$/He~{\scriptsize II}~$\lambda1640$ diagnostic. The two N~{\scriptsize V} emitters in GO 4287 are shown as magenta star (CEERS-1025, $z=8.7166$) and red diamond (CEERS-7902, $z=6.9827$, a LRD). We overplot the broad line AGN GS-3073 ($z=5.55$; open red pentagon) from \citet{Ji2024}. Open black squares show AGN from literature ($z\sim0-4$; \citealt{Kraemer2000,Baldwin2003,Kuraszkiewicz2004,Nagao2006b,Hainline2011,Matsuoka2011,Alexandroff2013,Dors2014}).}
\label{fig:nv_diag}
\end{figure}

The line ratios of resonant doublets (i.e., N~{\small V}, C~{\small IV}) can face significant radiative transfer effects, with the resulting line ratio depending on the column density and kinematics of the absorbing gas. 
The theoretical flux ratio of the doublet is $f_{{\rm NV}\lambda1239}/f_{{\rm NV}\lambda1243}=2$ \citep{Bickel1969}, assuming the level populations are dominated by collisional excitation. 
In the case that photon absorption dominates excitation, the doublet ratio becomes the ratio of Einstein A coefficients, yielding a flux ratio of $f_{{\rm NV}\lambda1239}/f_{{\rm NV}\lambda1243}=1$.
If we interpret the detected N~{\small V} emission feature as the N~{\small V}~$\lambda1243$, we can place a $3\sigma$ upper limit on the N~{\small V}~$\lambda1239$ flux ($<5.2\times10^{-19}$~erg~s$^{-1}$~cm$^{-2}$) and EW ($<2.7$~\AA). 
This would suggest the $f_{{\rm NV}\lambda1239}/f_{{\rm NV}\lambda1243}$ flux ratio in CEERS-1025 is $<0.37$ (at 3$\sigma$), very different from the theoretical value ($1-2$), regardless of whether collisions or absorption dominates excitation. 

One possible explanation for the observed doublet ratio could be the effect of N~{\small V} photons scattering through highly ionized outflowing gas.
In a simplified picture, the N~{\small V} emission is produced by ions near (or close to) the systemic redshift (similar to the other narrow lines in the spectrum), but the line profile is then altered by resonant scattering through a highly ionized outflowing medium. 
The N~{\small V} ions in the outflowing gas process a redshifted version of the spectrum. 
As a result, the N~{\small V}~$\lambda1243$ transition in the outflow will preferentially scatter photons on the blue side of the N~{\small V}~$\lambda1243$ emission line emitted by the galaxy. 
The precise outcome will depend on the geometry, column density, and velocity profile of the outflowing gas. 
If there is sufficient opacity extending over the velocity separation of the doublet ($\sim1000$~km~s$^{-1}$), this can lead to N~{\small V}~$\lambda1239$ photons being scattered away by the N~{\small V}~$\lambda1243$ transition in the outflowing gas. 
The  N~{\small V}~$\lambda1239$ photons will additionally likely face self-absorption from the N~{\small V}~$\lambda1239$ transition in the outflowing gas. 
In contrast, the N~{\small V}~$\lambda1243$ photons would only face scattering from the N~{\small V}~$\lambda1243$ transition, as there is no redder component. 
In this case, the $f_{{\rm NV}\lambda1239}/f_{{\rm NV}\lambda1243}$ doublet ratio would decrease relative to the theoretical value. 
In reality the situation may be more complicated if the outflowing gas is not uniformly distributed (e.g., \citealt{Wang2010}). 
Nevertheless, the kinematics of outflowing gas has been suggested to play a role in C~{\small IV} and Mg~{\small II} line profiles \citep[e.g.,][]{Chang2024,Topping2024} and may be impacting the N~{\small V} doublet in CEERS-1025.
Confirmation of this scenario would require a deeper spectrum capable of detecting the velocity profile of highly ionized gas in absorption, or integral field spectroscopy capable of detecting the diffuse flux from the scattered component of the N~{\small V} doublet. 
A higher S/N spectrum may also be expected to reveal broad emission in the detected N~{\small V}~$\lambda1243$ component if the outflow is scattering line photons up to $\sim1000$~km~s$^{-1}$.

Assuming the N~{\small V} emission in CEERS-1025 probes a hard radiation field, we may also expect to detect line emission from other highly-ionized species. 
The G140H observations cover N~{\small IV}], C~{\small IV}, He~{\small II}, O~{\small III}], N~{\small III}], but none of these features is detected (Figure~\ref{fig:CEERS-1025_spec}).  
The 3$\sigma$ limits on flux ratios between N~{\small V} and other UV lines are well above unity (N~{\small V}/N~{\small IV}] $>1.4$, N~{\small V}/C~{\small IV} $>1.4$, N~{\small V}/He~{\small II} $>2.6$). 
These flux ratios imply CEERS-1025 is a stronger nitrogen emitter than many AGN with narrow UV lines (where typically N~{\small V}/C~{\small IV} $<1$ and N~{\small V}/He~{\small II} $<2$, Figure~\ref{fig:nv_diag}; e.g., \citealt{Nagao2006b,Hainline2011,Alexandroff2013,Ji2024}). 
On the other hand, there are examples of AGN with similar N~{\small V}/C~{\small IV} ($\simeq1-10$) and N~{\small V}/He~{\small II} ratios ($\simeq2-10$; e.g., \citealt{Kraemer2000,Baldwin2003,Kuraszkiewicz2004,Jiang2008,Matsuoka2011,Dors2014}), many of which have been described as nitrogen-loud systems \citep[e.g.,][]{Baldwin2003,Jiang2008}. 
We note that the nitrogen-enhanced line ratios tend to refer to broad lines, whereas here we are observing similar line ratios from narrow line emitting gas. 
What drives the nitrogen excess in these AGN is not clear. 
In some cases, such large N~{\small V}/C~{\small IV} and N~{\small V}/He~{\small II} ratios have been argued as a signpost of supersolar metallicities \citep[e.g.,][]{Hamann1992,Baldwin2003,Nagao2006a}, whereas others have suggested that a nitrogen enhanced abundance pattern may contribute \citep[e.g.,][]{Ji2024}. 
We will discuss potential implications of the line ratios further in Section~\ref{sec:line_ratios}.

Finally, we do note that the N~{\small V} doublet is also ubiquitous in very hot stellar atmospheres, where it is most commonly encountered in P-Cygni profile prominent for O stars hotter than O6, corresponding to effective temperatures $\gtrsim 35$~kK \citep[e.g.,][]{Walborn1984}. 
It can also be produced in nearly pure emission in the hottest Wolf-Rayet atmospheres, where generally it is substantially broadened by the high velocities ($\gtrsim 1000$~km~s$^{-1}$) of the optically thick winds where it is produced \citep[e.g.,][]{Hainich2014}. 
While slower wind velocities are possible especially for very luminous/massive stars at extremely low metallicities \citep[e.g.,][]{Vink2022}, $208$~km~s$^{-1}$ (the FWHM of N~{\small V}~$\lambda1243$ of CEERS-1025) is extraordinarily low in the context of stellar winds. 
Crucially, it would also be atypical to encounter prominent N~{\small V} without any of the other strong wind features commonly encountered alongside it: most notably, C~{\small IV} and He~{\small II}, as well as N~{\small IV}]~$\lambda1486$ and N~{\small IV}~$\lambda1719$.
While we cannot entirely rule out a contribution from stellar winds, the requirements on these would be unusual.


\begin{figure}
\includegraphics[width=\linewidth]{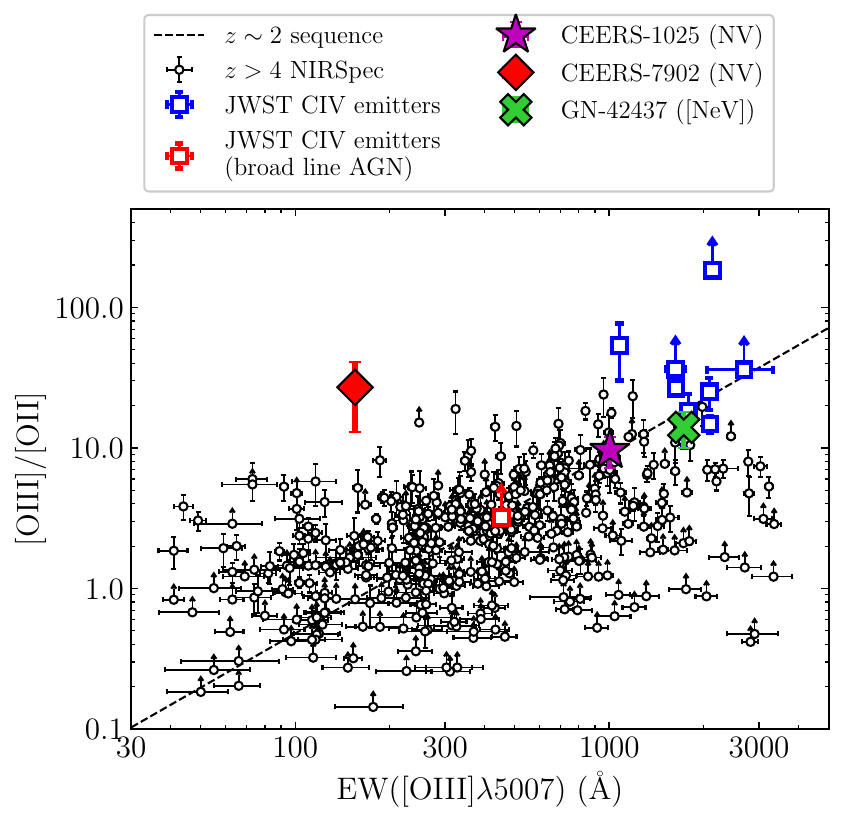}
\caption{[O~{\scriptsize III}]~$\lambda\lambda4959,5007$/[O~{\scriptsize II}]~$\lambda3727$ as a function of [O~{\scriptsize III}]~$\lambda5007$ EW. We show the two N~{\scriptsize V} emitters in GO 4287 CEERS-1025 as a magenta star and CEERS-7902 (LRD) as a red diamond. We overplot the [Ne~{\scriptsize V}]~$\lambda3427$ emitter GN-42437 in \citet{Chisholm2024} as green cross. We also overplot C~{\scriptsize IV} emitters in \citet{Topping2024} and \citet{Topping2025a} as open blue squares, where the broad-line AGN is shown as the open red square. The full $z>4$ spectroscopic sample is shown as open black circles. The relation derived from $z\sim2$ galaxies \citep{Tang2019} is shown as the black dashed line.}
\label{fig:o32_o3ew}
\end{figure}

\subsubsection{Rest-Frame Optical Emission Lines in CEERS-1025} \label{sec:CEERS-1025_opt}

The composite rest-frame optical spectrum of CEERS-1025 (bottom panel of Figure~\ref{fig:CEERS-1025_spec}) reveals a suite of strong lines ([O~{\small II}], [Ne~{\small III}], H$\beta$, [O~{\small III}]~$\lambda4959$, and [O~{\small III}]~$\lambda5007$), similar to that reported in the CEERS discovery spectrum (e.g., \citealt{Tang2023}). 
We additionally identify the tentative [O~{\small III}]~$\lambda4363$ auroral line (S/N $=2.5$). 
We note that the flux peak of [O~{\small III}]~$\lambda4363$ is slightly offset from that of H$\beta$ and [O~{\small III}]~$\lambda5007$ lines in the spatial direction in the 2D spectrum (1 pixel, $\simeq0.1$~arcsec), which could be due to the flux fluctuation.
We will discuss the [O~{\small III}]~$\lambda4363$ line below. 

As was the case in the original spectrum of CEERS-1025, no broad emission lines are detected, either in the Balmer lines or the forbidden oxygen lines. 
We can constrain the flux of broad H$\beta$ based on the non-detection.
Assuming a line width that is typical in broad-line AGN (FWHM $\simeq2000$~km~s$^{-1}$; e.g., \citealt{Harikane2023,Maiolino2024b,Matthee2024,Kocevski2025}; see also Section~\ref{sec:lrd}), the broad H$\beta$ line flux is $<5.0\times10^{-19}$~erg~s~cm$^{-2}$ at $3\sigma$. 
This indicates an upper limit on the broad-to-narrow H$\beta$ flux ratio of $<0.34$, well below that is typical of broad-line AGN ($\simeq1-2$; e.g., \citealt{Harikane2023,Kokorev2023,Ubler2023,Juodzbalis2024,Matthee2024}). 
Although we cannot fully rule out the presence of a broad emission lines, the existing spectrum suggests that CEERS-1025 is possibly a narrow line AGN with its broad line region obscured.  

The observed Balmer decrement is consistent with minimal attenuation. 
We measure H$\gamma$/H$\beta=0.47\pm0.15$, which is consistent with the intrinsic H$\gamma$/H$\beta$ ratio expected from case B recombination ($0.47$, assuming the appropriate electron temperature, see below paragraphs; \citealt{Osterbrock2006}). 
As noted above, the UV continuum slope of CEERS-1025 is very blue, with $\beta=-2.5$, further suggesting minimal dust attenuation, as is common in $z\gtrsim8$ galaxies. 
Based on these results, we do not correct the rest-frame optical emission lines for reddening.

\citet{Tang2023} concluded that the narrow rest-frame optical lines in CEERS-1025 appear broadly consistent with expectations for a galaxy dominated by a young stellar population ($2$~Myr assuming constant star formation history) formed in a recent burst with moderately low metallicity ([O~{\small III}]~$\lambda5007$/H$\beta$ flux ratio $=4.4\pm0.8$). 
The [O~{\small III}]+H$\beta$ EW of CEERS-1025 is large ($1535\pm66$~\AA), with a value among the upper $10\%$ of EWs observed at $z\simeq6-9$ \citep{Endsley2024}. 
Such large [O~{\small III}]+H$\beta$ EWs are also linked to the strong nebular C~{\small IV} emitters, thought to trace a population of metal poor massive star clusters (e.g., \citealt{Topping2025a}).
In star forming galaxies, we expect ionization-sensitive flux ratios (i.e. [O~{\small III}]/[O~{\small II}], hereafter O32) to increase with the rest-frame optical emission line EWs \citep{Tang2019,Sanders2020,Tang2023}. 
Assuming zero dust attenuation, we find a large O32 value ($=9.6\pm2.4$) that places CEERS-1025 on the same ionization vs. [O~{\small III}]+H$\beta$ EW sequence as star forming galaxies at high redshift (Figure~\ref{fig:o32_o3ew}). 
If an AGN is the source of the narrow N~{\small V} emission in CEERS-1025, it is not altering the strong rest-frame optical line ratios and EWs in a manner that can be easily distinguished from what is seen in normal star forming galaxies. 

\begin{figure}
\includegraphics[width=\linewidth]{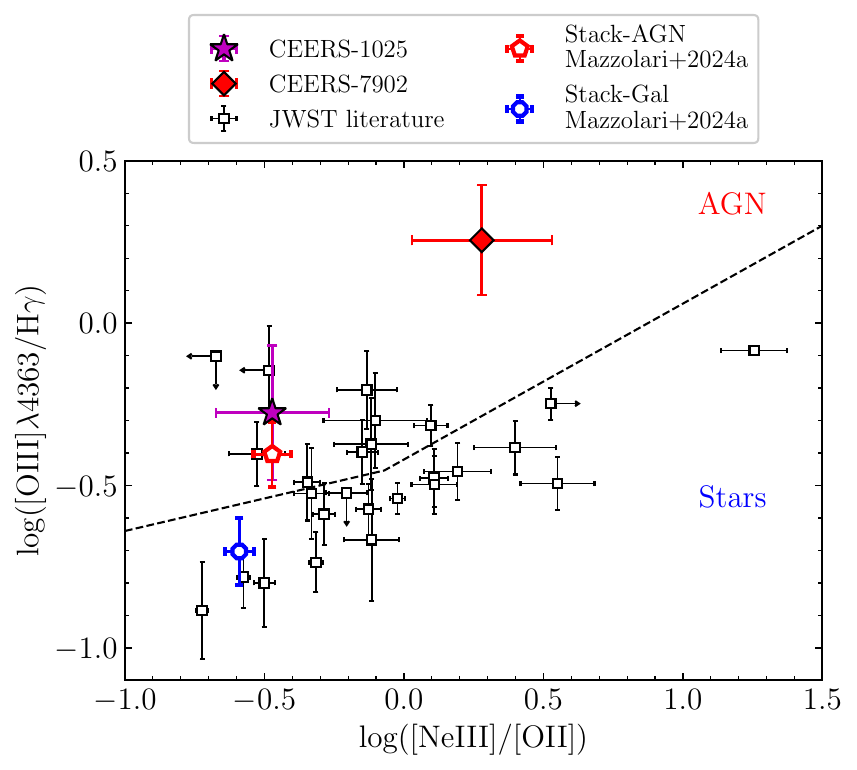}
\caption{[O~{\scriptsize III}]~$\lambda4363$/H$\gamma$ versus [Ne~{\scriptsize III}]~$\lambda3869$/[O~{\scriptsize II}] diagnostics, with black dashed line showing the demarcation between galaxies dominated by stars and AGN \citep{Mazzolari2024a}. We show the two N~{\scriptsize V} emitters CEERS-1025 as magenta star and CEERS-7902 (LRD) as red diamond. We overplot galaxies with [O~{\scriptsize III}]~$\lambda4363$ detections from literature {\it JWST} studies \citep{Nakajima2023,Hu2024,Sanders2024,Topping2024,Topping2025a} in open black squares, as well as the stacked narrow line AGN (open red pentagon) and stacked galaxies (open blue octagon) from \citep{Mazzolari2024b}.}
\label{fig:oiii4363_diag}
\end{figure}

An AGN power law spectrum should more efficiently heat the gas than a stellar ionizing spectrum, increasing the strength of auroral lines relative to hydrogen recombination lines \citep{Brinchmann2023,Mazzolari2024a,Ubler2024}. 
\citet{Mazzolari2024b} have developed a set of AGN diagnostics using a suite of photoionization models to derive a demarcation between AGN and stars using the combination of [O~{\small III}]~$\lambda4363$/H$\gamma$ and [Ne~{\small III}]/[O~{\small II}] (hereafter Ne3O2) flux ratios. 
The detection of a tentative [O~{\small III}]~$\lambda4363$ emission feature in the new composite G395M spectrum (Figure~\ref{fig:CEERS-1025_spec}) is suggestive of a large [O~{\small III}]~$\lambda4363$/H$\gamma$ ratio ($0.53\pm0.26$) that would place CEERS-1025 firmly in the AGN regime based on the \citet{Mazzolari2024b} diagnostics (Figure~\ref{fig:oiii4363_diag}). 
A deeper spectrum will be required to confirm this detection.

We can quantify the the physical properties that would be implied from the existing measurement of the auroral line strength. 
Using the \texttt{PYTHON} package \texttt{PyNeb} \citep{Luridiana2015} and assuming an electron density $n_{\rm e}\simeq1000$~cm$^{-3}$ that is typical at $z\sim9$ \citep[e.g.,][]{Isobe2023a}, we derive an electron temperature $T_{\rm e}$(O~{\small III}) $=3.0^{+1.3}_{-0.9}\times10^4$~K for the O$^{++}$ zone. 
Because the auroral [O~{\small II}]~$\lambda\lambda7320,7330$ lines are not covered by the NIRSpec spectra and are likely very faint, we derive the O$^+$ zone electron temperature $T_{\rm e}$(O~{\small II}) using the relation $T_{\rm e}$(O~{\small II}) $=0.7\times T_{\rm e}$(O~{\small III}) $+3000$~K \citep{Campbell1986,Garnett1992}.
Combining with the [O~{\small III}]~$\lambda5007$/H$\beta$ and [O~{\small II}]/H$\beta$ ratios, this indicates a gas-phase oxygen abundance $12+\log{\rm (O/H)}=7.18^{+0.24}_{-0.10}$ ($Z_{\rm neb}=0.03^{+0.02}_{-0.01}\ Z_{\odot}$, solar metallicity corresponds to $12+\log{\rm (O/H)}=8.71$; \citealt{Gutkin2016}). 
While the inferred electron temperature and oxygen abundance remains mostly unchanged over a wide electron density range $n_{\rm e}=10^2-10^4$~cm$^{-3}$, the calculated values would be different at yet higher electron densities seen in compact star forming galaxies (e.g., \citealt{Topping2024,Topping2025a}). 
At $n_{\rm e}\simeq10^5$~cm$^{-3}$, the inferred electron temperature decreases to $=2.4^{+0.9}_{-0.7}\times10^4$~K and the oxygen abundance becomes $12+\log{\rm (O/H)}=7.75^{+0.32}_{-0.24}$ ($Z_{\rm neb}=0.11^{+0.12}_{-0.05}\ Z_{\odot}$). 
While there is still considerable uncertainty in these measurements, the results suggest that the gas in CEERS-1025 is likely fairly metal poor.


\subsection{Spectroscopy of $z=6.99$ galaxy CEERS-7902} \label{sec:CEERS-7902}

CEERS-7902 (R.A. $=214.983038$, Decl. $=+52.956205$) is a broad line AGN that is relatively faint in the rest-frame UV continuum ($m_{\rm F115W}=27.0$, M$_{\rm UV}=-19.9$). 
The NIRCam photometry and initial characterization of this object was presented in \citet{Labbe2023}, identifying that CEERS-7902 (ID 38094 in their analysis) has a very red rest-frame optical SED ($\beta_{\rm{opt}}=0.89$; \citealt{Kocevski2025}), consistent with the LRD population. 
CEERS-7902 does not exhibit a significant UV upturn; instead its UV slope remains fairly red ($\beta_{\rm{UV}}=-0.75$) (see right panel of Figure~\ref{fig:SED}).

A low-resolution ($R\sim100$) NIRSpec prism spectrum and medium-resolution G395M/F290LP spectrum of CEERS-7902 was obtained in RUBIES. 
The RUBIES spectrum of CEERS-7902 (ID 55604 in RUBIES) confirmed the presence of broad H$\beta$ emission with narrow forbidden lines \citep{Wang2024,Kocevski2025}, leading to its characterization as a broad line AGN at $z=6.99$, as reported in \citet{Kocevski2025}. 
That paper also notes the presence of a narrow absorption line in the H$\beta$ profile (blueshifted from line center), as has been seen in $10-20\%$ of Type I AGN discovered with {\it JWST} \citep[e.g.,][]{Juodzbalis2024,Labbe2024,Lin2024,Matthee2024,Wang2024,Kocevski2025,Wang2025}. 
This may indicate the presence of neutral gas with sufficient density ($\gtrsim 10^8$ cm$^{-3}$) to populate the $n=2$ level of hydrogen. 
Using local scaling relations \citep{Greene2005}, \citet{Kocevski2025} infer a black hole mass of $M_{\rm BH}=2.0\times10^9\ M_{\odot}$ (after correcting for dust attenuation of the optical continuum). 
By jointly fitting AGN and stellar contributions to the SED, they infer a stellar mass of $M_{\star}=1.3\times10^{10}\ M_{\odot}$ and an optical attenuation of $A_V=3.4$, similar to the values reported in \citet{Wang2024}  ($M_{\star}=6.2\times10^9\ M_{\odot}$ and $A_V=3.8$).

\begin{figure*}
\includegraphics[width=\linewidth]{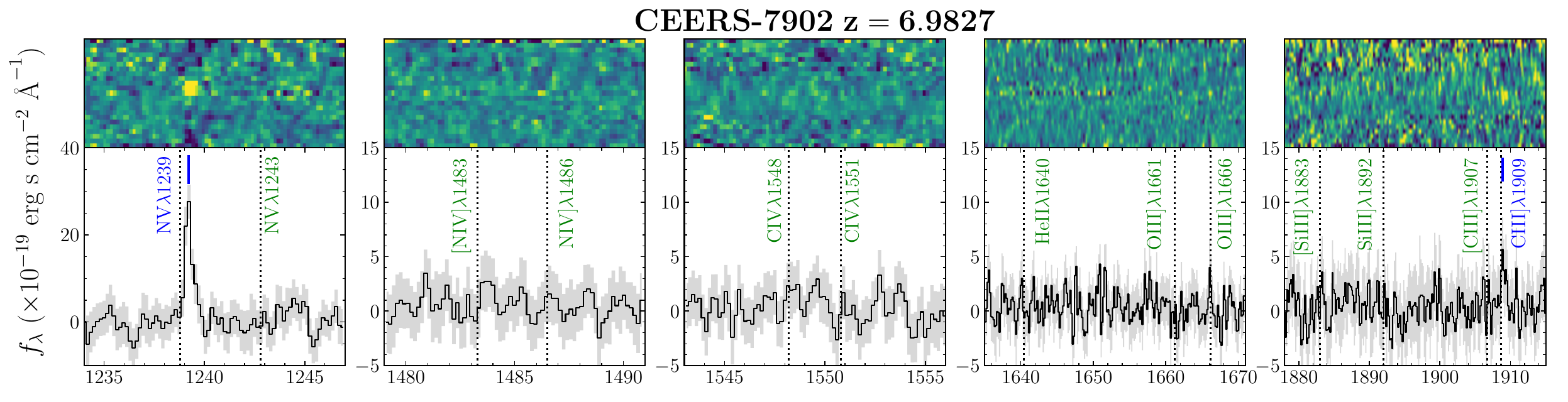}
\includegraphics[width=\linewidth]{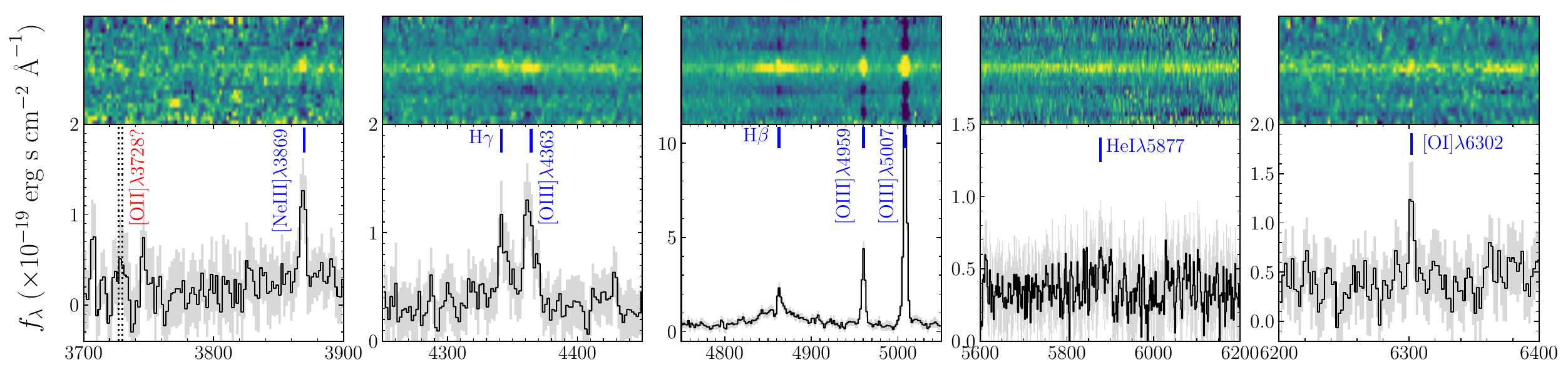}
\includegraphics[width=\linewidth]{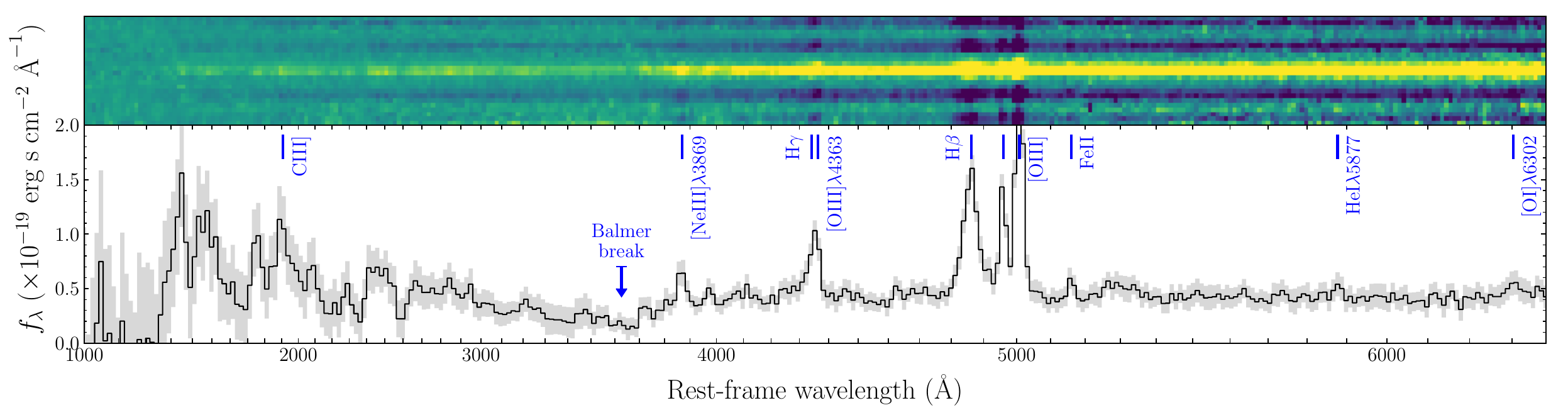}
\caption{NIRSpec spectra of the N~{\scriptsize V} emitter CEERS-7902 (LRD). We show the GO 4287 G140H, the composite of GO 4287 and RUBIES G395M, and the RUBIES prism spectra from top to bottom. Spectra are shown in the same way as Figure~\ref{fig:CEERS-1025_spec}. We find tentative detection of unresolved [O~{\scriptsize II}]~$\lambda3728$ emission line (S/N $=2$), which is marked by red text.}
\label{fig:CEERS-7902_spec}
\end{figure*}

Here we present the first deep rest-frame UV view of CEERS-7902 at high resolution using the G140H grating (Figure~\ref{fig:CEERS-7902_spec}) and a new moderate resolution G395M spectrum targeting the rest-frame optical. 
Below we describe the rest-frame UV emission line measurements (Section~\ref{sec:CEERS-7902_uv}). 
We then briefly comment on the rest-frame optical spectra of CEERS-7902 (Section~\ref{sec:CEERS-7902_opt}). 
In the following, we adopt a systemic redshift of $z_{\rm sys}=6.9827$ derived from the narrow [O~{\small III}] doublet and other strong forbidden lines in the composite G395M spectrum. 
This value is consistent with that measured for similar lines in the RUBIES grating spectrum of this galaxy.


\subsubsection{Rest-Frame UV Emission Lines in CEERS-7902} \label{sec:CEERS-7902_uv}

We visually search the G140H spectrum (top panel of Figure~\ref{fig:CEERS-7902_spec}) for emission lines, identifying a strong ($8\sigma$) feature at $9892.4$~\AA, close to the expected position of N~{\small V}~$\lambda1239$. 
We also find another confidently-detected ($3\sigma$) emission line at $15238.0$~\AA, consistent with the expected wavelength of the red component of the [C~{\small III}], C~{\small III}]~$\lambda\lambda1907,1909$ doublet. 
As described below, both detected  UV lines are narrow. Broad lines with FWHM similar to the broad H$\beta$ seen in the rest-frame optical are not detected in the  rest-frame UV spectrum of CEERS-7902.

The emission feature seen near the N~{\small V} resonance could either be interpreted as the blue component of the N~{\small V} doublet, with peak flux emerging slightly redshifted ($+104\pm39$~km~s$^{-1}$) from line center, or it could be interpreted as the red component of the N~{\small V} doublet with its peak flux significantly blueshifted ($-861\pm39$~km~s$^{-1}$). 
The observed line width of N~{\small V}~$\lambda1239$ emission (FWHM $=101\pm25$~km~s$^{-1}$) is similar to the instrument resolution ($\simeq111$~km~s$^{-1}$), suggesting that the N~{\small V}-emitting gas is kinematically distinct from the gas in the broad line region of the AGN.  
The measured flux ($1.20\pm0.15\times10^{-18}$~erg~s$^{-1}$~cm$^{-2}$) suggests an EW of $27.9\pm3.4$~\AA. 
Here we find the same EW value regardless of whether we use the stacked flux density in the spectrum or the NIRCam-measured rest-frame UV flux density underneath the emission line. 

We note that the detected feature near N~{\small V} is redshifted by $\simeq5800$~km~s$^{-1}$ from Ly$\alpha$. 
While it is conceivable that we are seeing Ly$\alpha$ that has been shifted in wavelength greatly owing to interaction with a very dense (and dust-free) column of hydrogen, we would  expect to see a broadened line profile in this case, not the narrow line in the spectrum. 
In what follows, we will assume that the detected feature in CEERS-7902 is nebular emission associated with the N~{\small V} doublet, likely implying photoionization from a source of $77$~eV photons. 
As noted in Section~\ref{sec:CEERS-1025_uv}, if the detected N~{\small V} feature was associated with stellar winds, we would expect to see other lines (i.e., C~{\small IV}, He~{\small II}, N~{\small IV}]) not present in the spectrum.

Only one of the N~{\small V} components is detected in the existing spectrum.  
We place a $3\sigma$ EW upper limit on the undetected component of $<7.7$~\AA. 
If we assume the observed line is N~{\small V}~$\lambda1239$, it would imply a doublet flux ratio limit of $f_{{\rm NV}\lambda1239}/f_{{\rm NV}\lambda1243}>3.7$ ($3\sigma$), much larger than the theoretically expected range of $1-2$. 
In the case that the detected feature is highly blueshifted emission from N~{\small V}~$\lambda1243$, it would imply a line ratio ($f_{{\rm NV}\lambda1239}/f_{{\rm NV}\lambda1243}<0.27$ at $3\sigma$) that is also inconsistent with theoretical values. 
As discussed in Section~\ref{sec:CEERS-1025_uv}, line ratios of resonant doublets can be impacted by scattering. 
To preferentially boost the blue component would require scattering through infalling material. 
This picture does not provide a fully satisfactory explanation of the line profile, as we will discuss in Section~\ref{sec:line_ratios}. 

The gas clouds responsible for the N~{\small V} emission appear to not emit strongly in C~{\small IV} or He~{\small II}. 
We place $3\sigma$ upper limits on the fluxes and EWs of N~{\small IV}], C~{\small IV}, He~{\small II}, O~{\small III}], and N~{\small III}]. 
The flux ratios between N~{\small V} and other UV lines are well above unity, with N~{\small V}/N~{\small IV}] $>2.4$, N~{\small V}/C~{\small IV} $>2.6$, and N~{\small V}/He~{\small II} $>4.8$. 
While these line ratios are not typical \citep[e.g.,][]{Nagao2006b,Alexandroff2013,Hainline2011,Ji2024}, they are consistent with those found in some populations of AGN, including many  nitrogen-loud AGN (e.g., \citealt{Kraemer2000,Baldwin2003,Kuraszkiewicz2004,Jiang2008,Matsuoka2011,Dors2014}; Figure~\ref{fig:nv_diag}, see also Section~\ref{sec:CEERS-1025_uv}).

The detection of the C~{\small III}]~$\lambda1909$ emission line provides further insight into the gas conditions associated with CEERS-7902. 
The line is narrow, with FWHM $=135\pm36$~km~s$^{-1}$, similar to the line width of N~{\small V} as well as the other narrow lines in the rest-frame optical.
The emission feature has a peak consistent with line center ($+28\pm26$~km~s$^{-1}$), using the systemic redshift implied by the forbidden lines in the rest-frame optical.
The measured flux ($3.5\pm1.2\times10^{-19}$~erg~s$^{-1}$~cm$^{-2}$) indicates a rest-frame EW of $22.5\pm7.7$~\AA. 
Such a large value (for an individual C~{\small III}] component) is rarely seen in star forming galaxies \citep[e.g.,][]{Shapley2003,Stark2014,Stark2015a,Hutchison2019,Du2020,Mainali2020,Tang2021,Roberts-Borsani2024} but is commonly exhibited by AGNs \citep{Nakajima2018,LeFevre2019}. 
The doublet flux ratio constrains the density of the C~{\small III}]-emitting gas. 
The absence of the the [C~{\small III}]~$\lambda$1907 line implies a limit of $f_{{\rm CIII]}\lambda1909}/f_{{\rm [CIII]}\lambda1907}>1.2$ at $3\sigma$. 
Using \texttt{PyNeb} and assuming an electron temperature $T_{\rm e}=2.4-10\times10^4$~K (see Section~\ref{sec:CEERS-7902_opt}), we derive an electron density of $n_{\rm e}>3.8\times10^4$~cm$^{-3}$ from the doublet ratio. 
Such high densities are not uncommon in gas traced by the C~{\small III}] doublet at high redshift \citep[e.g.,][]{Maiolino2024a,Senchyna2024,Topping2024,Topping2025a,Topping2025b}.

O~{\small I} emission lines are often seen in in AGN spectra \citep[e.g.,][]{Grandi1980,Rodriguez-Ardila2002,Riffel2006,Martinez-Aldama2015,Cracco2016,Juodzbalis2024,Wang2025}. 
The strength of three of these lines ($1304$, $8446$, and $11287$~\AA) can be boosted significantly by Ly$\beta$ pumping, owing to the similarity of the Ly$\beta$ resonance wavelength and that of the 3d$^3$ D$^0$ excited state of the O~{\small I} atom \citep{Kwan1981}, the decay of which produces O~{\small I} photons with the aforementioned wavelengths. 
Several {\it JWST}-detected AGNs have shown the two reddest O~{\small I} lines, but confirmation of the Ly$\beta$ fluorescence picture requires detection of the $1304$~\AA\ line. 
CEERS-7902 should be a candidate for Ly$\beta$ pumping of the O~{\small I} transitions owing to the indications of extremely dense gas from H$\beta$ absorption (and potentially the Balmer break). 
However we do not see the  O~{\small I}~$\lambda1304$ in the spectrum, and the two redder transitions associated with Ly$\beta$ pumping are not covered by existing spectroscopy. 
We place a $3\sigma$ upper limit of $3.1\times10^{-19}$~erg~s$^{-1}$~cm$^{-2}$ on the line flux of O~{\small I}~$\lambda1304$. 
The absence of the line is likely related in part to the strong reddening that is present in the rest-frame UV of CEERS-7902. 
A considerably deeper spectrum would be required to usefully test the existence of the O~{\small I} transitions. 

\subsubsection{Rest-Frame Optical Emission Lines in CEERS-7902} \label{sec:CEERS-7902_opt}

The composite G395M spectrum of CEERS-7902 (middle panel of Figure~\ref{fig:CEERS-7902_spec}) reveals a suite of rest-frame optical emission lines ([Ne~{\small III}], H$\gamma$, H$\beta$, and [O~{\small III}]). 
The H$\beta$ line shows a broad emission component with a narrow absorption feature \citep{Wang2024,Kocevski2025}.
Broad emission is also seen in He~{\small I}~$\lambda5877$ in the G395 spectrum.
We additionally detect narrow emission at the expected wavelengths of [O~{\small III}]~$\lambda$4363 (S/N $=5$), and we  detect unresolved [O~{\small II}]~$\lambda3728$ doublet (S/N $=2$) as well as [O~{\small I}]~$\lambda6302$ (S/N $=4$) in the G395 spectrum. 
We also detect a tentative emission near [Fe~{\small VII}]~$\lambda5159$ (S/N $=1.5$).

We simultaneously fit the narrow and broad H$\beta$ emission profile with two Gaussians. 
We derive FWHMs of $274\pm40$~km~s$^{-1}$ (narrow) and $3560\pm210$~km~s$^{-1}$ (broad). 
We find a similar FWHM for the broad He I $\lambda5877$ component ($2731\pm635$~km~s$^{-1}$).
The spectral width of the broad H$\beta$ component is consistent to that measured in \citet{Wang2024} from RUBIES G395M spectrum (FWHM $=3595\pm250$~km~s$^{-1}$). 
It is lower than that presented in \citet{Kocevski2025} (FWHM $=4870\pm480$~km~s$^{-1}$). 
A higher S/N spectrum should yield a more robust measure, but for the purposes of this paper, the precise width of the broad component is not critical. 
We also characterize the H$\beta$ absorption feature reported in the literature \citep{Wang2024,Kocevski2025}. 
The absorption is strong (EW $=-3.3$~\AA), narrow (FWHM $=334$~km~s$^{-1}$), and slightly blueshifted with respect to the line center ($-321$~km~s$^{-1}$). 
The rest-frame optical spectrum of CEERS-7902 also shows a Balmer break (bottom panel of Figure~\ref{fig:CEERS-7902_spec}) which could either be due to stars \citep{Wang2024} or Balmer limit absorption from dense gas \citep[e.g.,][]{Inayoshi2025,Ji2025}.

Interpretation of narrow lines requires insight of whether they are significantly reddened by dust. 
The rest-frame optical spectrum of CEERS-7902 shows detection of narrow H$\gamma$ and H$\beta$ emission, allowing us to measure the Balmer decrement of the narrow line emitting gas.
We simultaneously fit the narrow and broad H$\gamma$ emission as well as [O~{\small III}]~$\lambda4363$ with three Gaussians, fixing the line width of the broad H$\gamma$ to the value derived from broad H$\beta$. 
From this, we find that the narrow line Balmer decrement is H$\gamma$/H$\beta=0.67\pm0.27$. 
If there was significant reddening of the narrow lines, we would expect the ratio to be significantly lower than the intrinsic ratio expected from case B recombination ($0.47$; \citealt{Osterbrock2006}). 
The fact that we measure a value that is larger than intrinsic (albeit consistent within $1\sigma$) gives no strong evidence that the narrow line emitting gas is strongly impacted by dust, although we note there may be deviations from case B \citep[e.g.,][]{Scarlata2024,McClymont2025}. 
In the following, we will not correct the narrow lines for reddening. 

The narrow [O~{\small III}]$\lambda5007$ line is the strongest line in the spectrum  ($5.77\pm0.17\times10^{-18}$~erg~s$^{-1}$~cm$^{-2}$), but the derived rest-frame optical [O~{\small III}]~$\lambda5007$ EW ($=155\pm5$~\AA) is well below average at $z\simeq6-9$ \citep[e.g.,][]{Endsley2024}. 
The line ratio between [O~{\small III}]~$\lambda5007$ and the narrow component of H$\beta$ in CEERS-7902 is $9.8\pm1.9$, larger than typically seen in star-forming galaxies at similar redshifts \citep[e.g.,][]{Sanders2023,Shapley2023,Shapley2025,Trump2023,Backhaus2024,Kumari2024}, 
The ionization-sensitive flux ratios are also large. 
The Ne3O2 ratio ($=1.9\pm1.1$) of CEERS-7902 is larger than that of the vast majority of galaxies at high redshift \citep[e.g.,][]{Cameron2023b,Nakajima2023,Shapley2023,Tang2023,Trump2023,Backhaus2024,Hu2024,Sanders2024,Shapley2025}. 
The O32 ratio ($=27\pm14$) also points to gas under extreme ionization conditions. 
Such large Ne3O2 and O32 ratios are rarely seen in galaxies with as low  [O~{\small III}] EW as is seen in CEERS-7902 (Figure~\ref{fig:o32_o3ew}). 
This likely reflects the additional contribution of the AGN continuum to the rest-frame optical, reducing the [O~{\small III}] EW relative to systems dominated by stellar continuum. 


\begin{figure*}
\includegraphics[width=\linewidth]{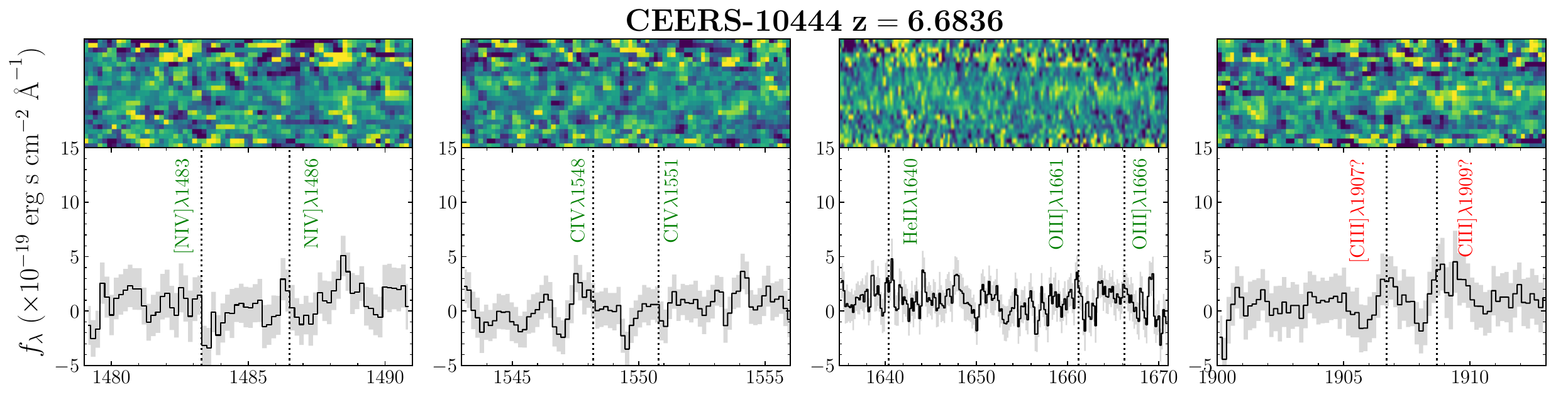}
\includegraphics[width=\linewidth]{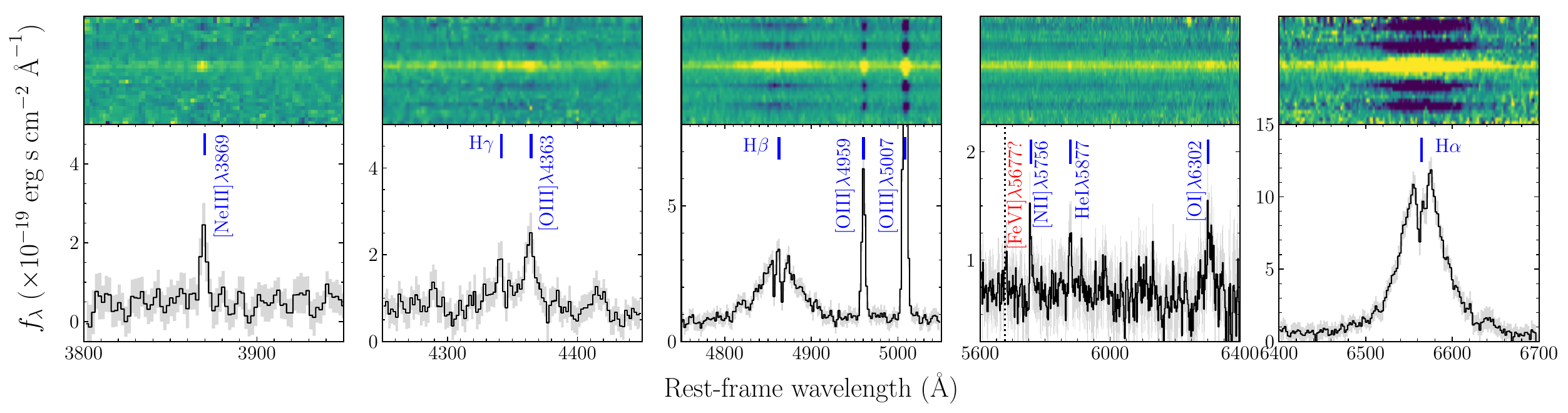}
\caption{NIRSpec spectra of CEERS-10444 (LRD). We show the GO 4287 G140H spectrum in the top panel and the composite G395M spectrum (stacking the GO 4287 and RUBIES grating spectra) in the bottom panel. Spectra are shown in the same way as Figure~\ref{fig:CEERS-1025_spec}.}
\label{fig:CEERS-10444_spec}
\end{figure*}

The detection of [O~{\small III}]~$\lambda4363$ feature allows us to apply the diagnostics presented in \citet{Mazzolari2024a} to the narrow line gas in CEERS-7902. 
The measured narrow line [O~{\small III}]~$\lambda4363$/H$\gamma$ ratio ($1.8\pm0.7$) is much larger than the limits implied by other galaxies in GO 4287 and by other galaxies with [O~{\small III}]~$\lambda4363$ detections in earlier {\it JWST} programs \citep[e.g.,][]{Nakajima2023,Hu2024,Mazzolari2024b,Sanders2024,Topping2024,Topping2025a}. 
The [O~{\small III}]~$\lambda4363$/H$\gamma$ and [Ne~{\small III}]/[O~{\small II}] line ratios place CEERS-7902 in the regime expected for AGN in the models presented in \citet{Mazzolari2024a} (Figure~\ref{fig:oiii4363_diag}). 
Although galaxies dominated by massive stars with large densities ($n_{\rm e}\gtrsim10^5$~cm$^{-3}$) could also have high [O~{\small III}]~$\lambda4363$/H$\gamma$ ratios, such large densities will result in weaker [O~{\small II}], resulting in very large Ne3O2 and O32 ratios \citep[e.g.,][]{Topping2024,Topping2025a}. 
The combination of large [O~{\small III}]~$\lambda4363$/H$\gamma$ and relatively low O32 and Ne3O2 in CEERS-7902 is consistent with expectations for AGN photoionization. 
We also infer the electron temperature from the measured [O~{\small III}]~$\lambda5007$/[O~{\small III}]~$\lambda4363$ ratio of CEERS-7902 ($7.9\pm1.5$). 
Using \texttt{PyNeb} and assuming an electron density varying between $3.8\times10^4$~cm$^{-3}$ (as computed from the lower limit of the [C~{\small III}], C~{\small III}] doublet flux ratio) and $6.4\times10^5$~cm$^{-3}$ (the critical density of [O~{\small III}]~$\lambda4363$), we derive an electron temperature in the range between $2.4\times10^4$~K and $1\times10^5$~K.

The tentative [O~{\small I}]~$\lambda6302$ emission line may also point to the presence of AGN photoionization.
Because H$\alpha$ is shifted out of the G395M spectrum, we estimate the H$\alpha$ flux from the observed narrow H$\beta$ flux assuming the case B value, H$\alpha$/H$\beta=2.76$ \citep{Osterbrock2006}. 
The estimated narrow H$\alpha$ line flux is $1.62\times10^{-18}$~erg~s$^{-1}$~cm$^{-2}$. 
This indicates an [O~{\small I}]~$\lambda6302$/H$\alpha$ ratio $=0.27\pm0.09$ for the narrow line emitting gas.
Together with the [O~{\small III}]~$\lambda5007$/H$\beta$ ratio ($=9.8$ for the NLR), these line ratios suggest the the gas in CEERS-7902 that emits in narrow lines is in the AGN regime of the diagnostics in \citet{Mazzolari2024b}. 
The [O~{\small I}]~$\lambda6302$/H$\alpha$ ratio is $\simeq10\times$ larger than the typical value of stars at fixed [O~{\small III}]~$\lambda5007$/H$\beta$ ratio.  

\subsection{Spectroscopy of $z=6.68$ galaxy CEERS-10444} \label{sec:CEERS-10444}

We present the rest-frame UV and optical spectra of CEERS-10444 (R.A. $=214.892231$, Decl. $=+52.877651$). 
This system is relatively faint in the rest-frame UV continuum ($m_{\rm F115W}=26.9$, M$_{\rm UV}=-19.9$). 
The NIRCam photometry of CEERS-10444 was presented in \citet{Kocevski2025}, identifying that it has a red rest-frame optical SED ($\beta_{\rm opt}=0.43$) that is consistent with the LRD population. 
Both a NIRSpec prism spectrum and NIRSpec G395M/F290LP spectrum were obtained in RUBIES (ID 49140 in RUBIES) and discussed in \citet{Kocevski2025} and \citet{Wang2024}. 
The RUBIES spectra show broad H$\alpha$ and H$\beta$ emission as well as narrow forbidden lines, leading to its characterization as a broad line AGN at $z=6.68$. 
Similar to CEERS-7902, narrow absorption lines are also present in the H$\alpha$ and H$\beta$ profiles of CEERS-10444 \citep{Wang2024,Kocevski2025}, likely indicating the presence of dense neutral gas populating the $n=2$ level of hydrogen \citep{Inayoshi2025}.
Balmer break is shown in the RUBIES prism spectrum of CEERS-10444.
We measure a Balmer break strength $f_{\nu,4050}/f_{\nu,3670}=3.4\pm0.5$, which is consistent with that measured in \citet{Wang2024} within $2\sigma$ ($2.44\pm0.10$).

In the top panel of Figure~\ref{fig:CEERS-10444_spec}, we present the first deep, high-resolution rest-frame UV spectrum of CEERS-10444 from GO 4287 with G140H.
We also generate a composite G395M spectrum of CEERS-10444 (bottom panel of Figure~\ref{fig:CEERS-10444_spec}) by stacking the GO 4287 and RUBIES G395M spectra. 
Using the narrow [O~{\small III}] doublet and other strong forbidden lines in the composite G395M spectrum, we derive a systemic redshift of $z_{\rm sys}=6.6836$. 
This is consistent with the redshift measured using the RUBIES grating spectrum \citep{Wang2024,Kocevski2025}. 
In this subsection we will adopt this systemic redshift for CEERS-10444.

\subsubsection{Rest-Frame UV Emission Lines in CEERS-10444} \label{sec:CEERS-10444_uv}

The G140H spectrum covers between $0.98$ and $1.80\ \mu$m, corresponding to $1275$ to $2345$~\AA\ in the rest-frame (top panel of Figure~\ref{fig:CEERS-10444_spec}).
We visually search for rest-frame UV emission lines in the G140H spectrum.
We tentatively identify two emission lines at $14651.7$~\AA\ ($2\sigma$) and $14667.3$~\AA\ ($3\sigma$), respectively. 
They are consistent with the expected wavelengths of the [C~{\small III}], C~{\small III}]~$\lambda\lambda1907,1909$ doublet, with the peak of the blue (red) emission component $+28\pm27$~km~s$^{-1}$ ($+33\pm27$~km~s$^{-1}$) away from the line center of [C~{\small III}]~$\lambda1907$ (C~{\small III}]~$\lambda1909$). 
Both lines are narrow, with FWHM $=127\pm27$~km~s$^{-1}$ for [C~{\small III}]~$\lambda1907$ and FWHM $=214\pm27$~km~s$^{-1}$ for C~{\small III}]~$\lambda1909$, similar to that of narrow forbidden lines in rest-frame optical. 
The measured fluxes ($f_{{\rm [CIII]}\lambda1907}=1.6\pm0.8\times10^{-19}$~erg~s$^{-1}$~cm$^{-2}$, $f_{{\rm CIII]}\lambda1909}=4.0\pm1.2\times10^{-19}$~erg~s$^{-1}$~cm$^{-2}$) indicate rest-frame EW $=4.8\pm2.4$~\AA\ for [C~{\small III}]~$\lambda1907$ and $=11.6\pm3.6$~\AA\ for C~{\small III}]~$\lambda1909$. 
Using the doublet flux ratio ($f_{{\rm CIII]}\lambda1909}/f_{{\rm [CIII]}\lambda1907}=2.4\pm1.4$), we can constrain the density of the C~{\small III}]-emitting gas. 
Using \texttt{PyNeb} and assuming an electron temperature $T_{\rm e}=5.5\times10^4$~K (see Section~\ref{sec:CEERS-10444_opt}), we derive a large electron density $n_{\rm e}=1.6^{+1.3}_{-1.1}\times10^5$~cm$^{-3}$. 

We do not detect either C~{\small IV} or He~{\small II} in the G140H spectrum of CEERS-7902. 
We place $3\sigma$ upper limits on the EWs of EW$_{{\rm CIV}\lambda1548}<5.8$~\AA, EW$_{{\rm CIV}\lambda1551}<6.6$~\AA, and EW$_{{\rm HeII}\lambda1640}<6.4$~\AA. 
The N~{\small V} doublet is not covered by the G140H spectrum so we cannot verify if the line is present. 
We verify that the other nitrogen-based lines in the rest-frame UV ([N~{\small III}] and N~{\small IV}]) are not present in the spectrum of CEERS-10444, with $3\sigma$ upper limits of $<8.6$ and $<8.0$~\AA. 
We note that the tentative detection of C~{\small III}] and non-detection of C~{\small IV} and He~{\small II} is consistent with what was seen in CEERS-7902, so it is plausible that both LRDs have similar rest-frame UV spectra. 
A bluer spectrum is required to test this possibility, verifying the strength of the N~{\small V} doublet.

\subsubsection{Rest-Frame Optical Emission Lines in CEERS-10444} \label{sec:CEERS-10444_opt}

The composite G395M spectrum of CEERS-10444 (bottom panel of Figure~\ref{fig:CEERS-10444_spec}) reveals a suite of strong rest-frame optical emission lines ([Ne~{\small III}], H$\gamma$, H$\beta$, [O~{\small III}], H$\alpha$). 
Both H$\beta$ and H$\alpha$ show broad emission components with absorption features (see also \citealt{Wang2024,Kocevski2025}).
Broad emission is also seen in H$\gamma$.
We additionally detect narrow [O~{\small III}]~$\lambda4363$ emission (S/N $=5$), narrow [N~{\small II}]~$\lambda5756$ emission (S/N $=4$), and narrow He~{\small I}~$\lambda5877$ emission (S/N $=4$).
We detect a slightly broadened [O~{\small I}]~$\lambda6302$ emission line (S/N $=5$).
We also report several tentative iron emission lines (S/N $\simeq2$). We detect tentative emission near [Fe~{\small II}]~$\lambda4288$, blended O~{\small II}~$\lambda4416$+Fe~{\small II}~$\lambda4418$, [Fe~{\small VII}]~$\lambda5159$, Fe~{\small II}~$\lambda5199$, and [Fe~{\small VI}]~$\lambda5677$, which are often found in type I AGN \citep[e.g.,][]{Dong2010,Rose2015}.
The [O~{\small II}] doublet is shifted out of the blue end of the G395M spectrum.

The H$\beta$ emission of CEERS-10444 shows a broad component and two absorption components. 
We simultaneously fit the broad and the two absorption profiles of H$\beta$ with three Gaussians. 
For the broad component we derive FWHM $=3203\pm112$~km~s$^{-1}$, which is consistent to that measured in \citet{Wang2024} with the RUBIES G395M spectrum (FWHM $=3301\pm173$~km~s$^{-1}$). 
The two H$\beta$ absorption features are blueward ($-301$~km~s$^{-1}$) and redward ($+226$~km~s$^{-1}$) the systemic redshift, respectively.
The blueshifted H$\beta$ absorption is strong (EW $=-6.6$~\AA) and narrow (FWHM $=68$~km~s$^{-1}$).
The redshifted H$\beta$ absorption is even stronger (EW $=-8.2$~\AA) with a slightly wider profile (FWHM $=213$~km~s$^{-1}$).
We note that it is possible that this is a single absorption feature near line center (similar to that presented in \citet{Naidu2025} with a narrow emission line filling in the center ($v=-61\pm116$~km~s$^{-1}$ away from the line center), creating the appearance of two absorption features. 

The H$\alpha$ profile of CEERS-10444 appears mostly similar to H$\beta$, with a broad emission component and strong absorption. 
We simultaneously fit the H$\alpha$ profile with two Gaussians.
For the broad component we derive FWHM $=2686\pm39$~km~s$^{-1}$, roughly consistent with the line width of the broad H$\beta$ emission.
The H$\alpha$ absorption is strong (EW $=-30$~\AA), narrow (FWHM $=244$~km~s$^{-1}$), and blueshifted to the line center ($-109$~km~s$^{-1}$).
There is potentially a faint narrow emission line near line center ($v=+73\pm86$~km~s$^{-1}$) that has been almost entirely filled in, similar to the central emission feature seen in H$\beta$. 
A higher resolution and higher S/N spectrum would help better characterize the line profile. 

The H$\gamma$ of CEERS-10444 shows a broad and a narrow emission components. 
The broad component is blended with the auroral [O~{\small III}]~$\lambda4363$ line. 
We simultaneously fit the broad and narrow H$\gamma$ as well as [O~{\small III}]~$\lambda4363$ profiles with three Gaussians.
The broad H$\gamma$ has a line width (FWHM $=4169\pm613$~km~s$^{-1}$) consistent with that of H$\beta$.
The measured narrow line [O~{\small III}]~$\lambda4363$/H$\gamma$ ratio ($2.8\pm1.3$) is very large, consistent with many AGNs (see Figure~\ref{fig:oiii4363_diag}). 
The detection of [O~{\small III}]~$\lambda4363$ allows us to infer the electron temperature. 
The measured [O~{\small III}]~$\lambda5007$/[O~{\small III}]~$\lambda4363$ ratio of CEERS-10444 is $7.8\pm1.7$. 
Using \texttt{PyNeb} and assuming the electron density $n_{\rm e}=1.6\times10^5$~cm$^{-3}$ (as computed from the [C~{\small III}], C~{\small III}] doublet flux ratio), we derive an electron temperature of $T_{\rm e}=5.5^{+5.5}_{-2.0}\times10^4$~K.

The [N~{\small II}]~$\lambda5755$ auroral line is narrow (FWHM $=319\pm60$~km~s$^{-1}$).
As [N~{\small II}]~$\lambda5755$ is usually faint, the detection of [N~{\small II}]~$\lambda5755$ in the spectrum of CEERS-10444 (EW $=6.5\pm1.6$~\AA) may suggest a high electron temperature in the low-ionization gas. 
Unfortunately [N~{\small II}]~$\lambda6584$ emission is blended with the broad, strong H$\alpha$ emission, preventing us to compute a temperature based on the [N~{\small II}]~$\lambda5755$/[N~{\small II}]~$\lambda6584$ ratio. 
The He~{\small I}~$\lambda5877$ emission is relatively narrow (FWHM $=439\pm87$~km~s$^{-1}$) compared to broad Balmer emission lines. 
As the line is permitted, we expect there to be a broad component similar to that seen in CEERS-7902 (see Section~\ref{sec:CEERS-7902_opt}). 
It is likely that the broad He~{\small I} emission has a low S/N and thus is not clearly detected in the spectrum. 

The [O~{\small I}]~$\lambda6302$ emission line is strong, as is commonly expected in the partially ionized regions of AGN. 
For the other LRD in our sample, CEERS-7902, we demonstrated that the [O~{\small I}]~$\lambda6302$/H$\alpha$ ratio is consistent with that seen in many AGN spectra. 
We cannot measure the  [O~{\small I}]~$\lambda6302$/H$\alpha$ ratio for CEERS-10444 since we do not detect the narrow component of the Balmer lines. 
We do note that the  [O~{\small I}]~$\lambda6302$ actually appears somewhat broader than the narrow forbidden lines ([O~{\small III}], [N~{\small II}]), with FWHM $=1132\pm176$~km~s$^{-1}$. 
The [O~{\small I}] line is narrower than the broad Balmer emission lines, as we expect since the critical density of [O~{\small I}]~$\lambda6302$ is not large enough for the line to trace the dense broad line region. 
Deeper data are required to confirm the broadening of [O~{\small I}]~$\lambda6302$ and its physical origin. 
 
The [Fe~{\small VI}]~$\lambda\lambda5146, 5677$ doublet is sensitive to the electron temperature. 
While we tentatively detect the red component of the doublet, we do not detect the blue component. 
We place a $3\sigma$ lower limit on the [Fe~{\small VI}]~$\lambda5677$/[Fe~{\small VI}]~$\lambda5146$ line ratio $>0.74$. 
Using \texttt{PyNeb} and assuming the electron density $n_{\rm e}=1.6\times10^5$~cm$^{-3}$ inferred from [C~{\small III}], C~{\small III}] doublet flux ratio, we find that the observed [Fe~{\small VI}]~$\lambda5677$/[Fe~{\small VI}]~$\lambda5146$ ratio is consistent with that is expected ($0.86$) from the electron temperature inferred from [O~{\small III}]~$\lambda4363$ ($T_{\rm e}=5.5\times10^4$~K).

To summarize, CEERS-10444 is an LRD with a rest-frame optical spectrum that is similar in some respects to CEERS-7902, with both having broad hydrogen Balmer lines with strong absorption features. 
The rest-frame UV spectrum of CEERS-10444 shows [C~{\small III}], C~{\small III}] but not C~{\small IV} or He~{\small II}, as is also seen in CEERS-7902. 
However our G140H spectrum does not extend down to rest-frame $1240$~\AA\ in CEERS-10444, so we cannot verify if this source powers strong N~{\small V} emission.

\begin{figure*}
\includegraphics[width=\linewidth]{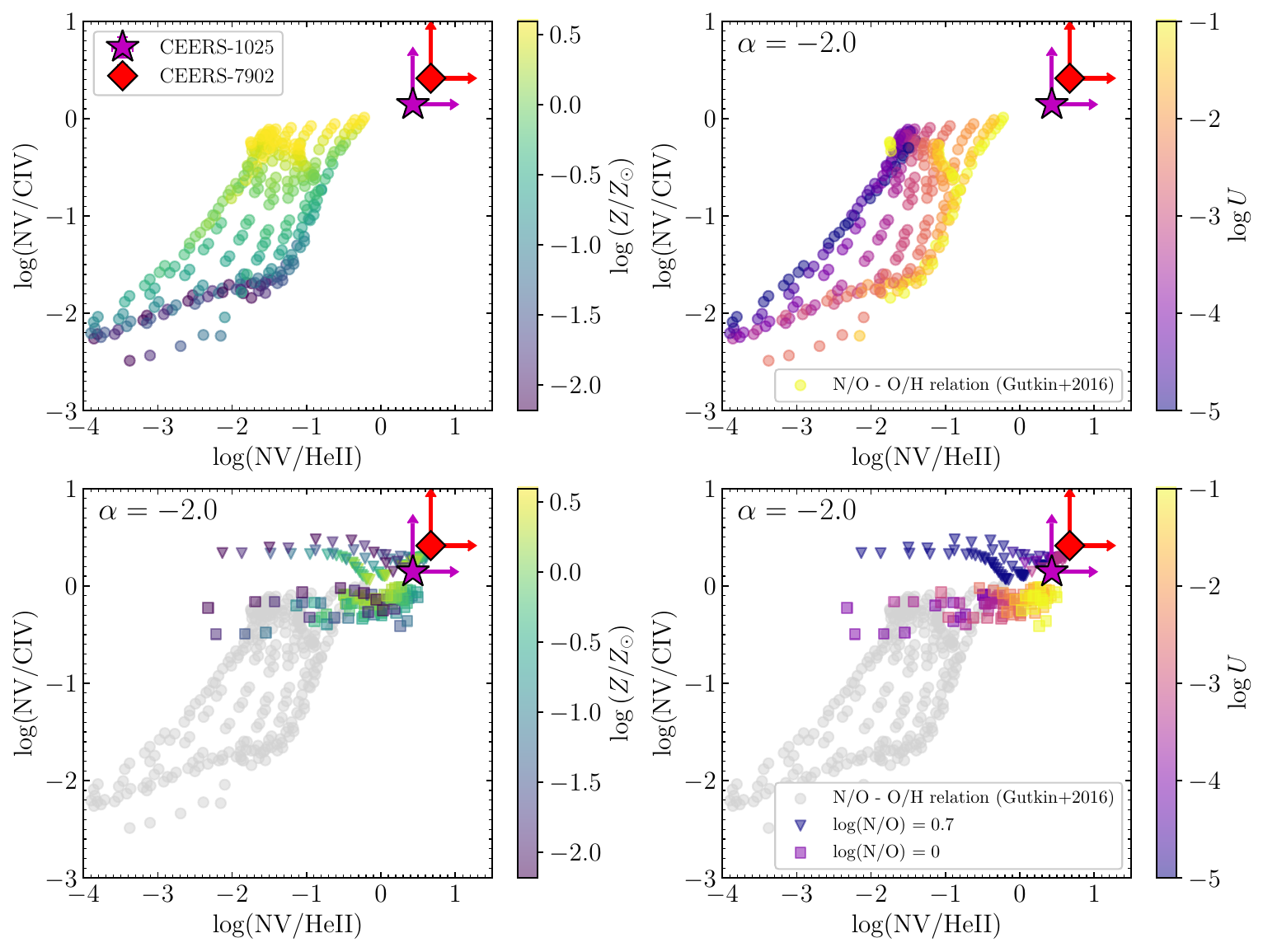}
\caption{N~{\scriptsize V}/C~{\scriptsize IV} versus N~{\scriptsize V}/He~{\scriptsize II} line ratios of the two N~{\scriptsize V} emitters CEERS-1025 (purple star) and CEERS-7902 (red diamond) and line ratios expected from photoionization models of narrow line AGN. The emitted spectrum of AGN accretion disk is assumed to be a power law $f_{\nu}\propto\nu^{\alpha}$ between $\lambda=0.001\ \mu$m to $\lambda=0.25\ \mu$m with $\alpha=-2.0$. The left two panels show models with different metallicity ($\log{(Z/Z_{\odot})}=-2.2$ to $0.6$) and the right two panels show models with different ionization parameters ($\log{U}=-5.0$ to $-1.0$). Models assuming N/O increasing with O/H following the prescription of \citet{Gutkin2016} are shown as circles. Models assuming fixed N/O are shown as upside-down triangles ($\log{\rm (N/O)}=0.7$) and squares ($\log{\rm (N/O)}=0$).}
\label{fig:UV_diag_alpha2p0}
\end{figure*}

\begin{figure*}
\includegraphics[width=\linewidth]{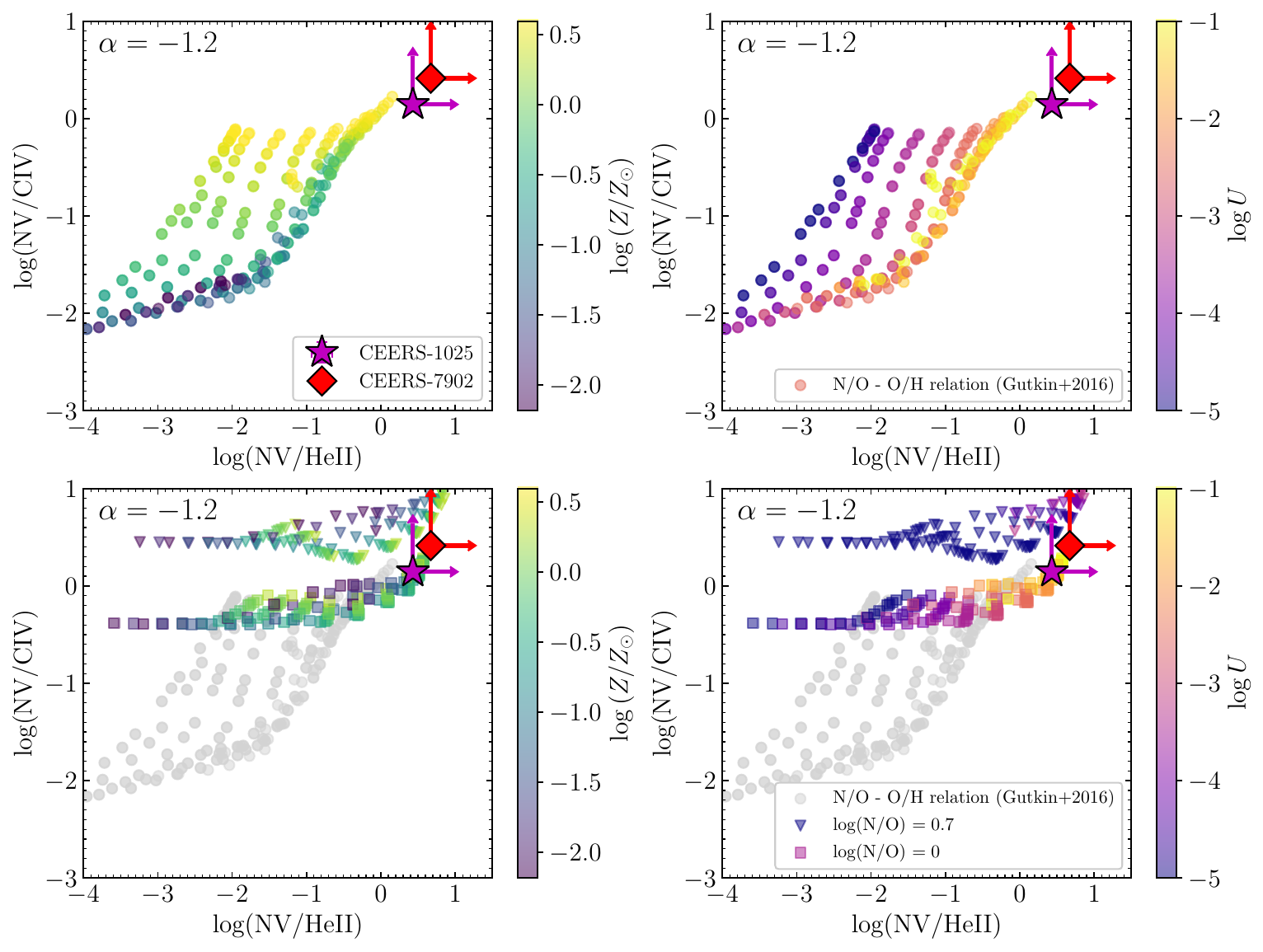}
\caption{N~{\scriptsize V}/C~{\scriptsize IV} versus N~{\scriptsize V}/He~{\scriptsize II} line ratios of CEERS-1025 (purple star) and CEERS-7902 (red diamond), and line ratios expected from photoionization models with AGN accretion disk power law spectrum $\alpha=-1.2$. Symbols are shown in the same way as Figure~\ref{fig:UV_diag_alpha2p0}.}
\label{fig:UV_diag_alpha1p2}
\end{figure*}

\subsection{Interpretation of Line Ratios} \label{sec:line_ratios}

We have detected the likely presence of narrow N~{\small V} emission in two of the galaxies observed as part of GO 4287,  CEERS-1025 and CEERS-7902. 
Both systems are undetected in C~{\small IV} and He~{\small II}, with limits indicating elevated N~{\small V}/C~{\small IV} and N~{\small V}/He~{\small II} ratios. 
One of the two galaxies (CEERS-7902) is a known broad line AGN with an SED consistent with most LRD selections. 
In this section, we consider what the line ratios may be revealing about these sources. 
We will focus on the highest ionization lines in the new rest-frame UV spectra as those are most likely to not have contribution from stars. 
Our goal in this section is primarily to investigate the range of factors that might adjust the N~{\small V}-based line ratios and not to derive the specific properties of the two N~{\small V} emitters.

We first consider whether the N~{\small V}/He~{\small II} and N~{\small V}/C~{\small IV} line ratios are plausibly consistent with expectations for photoionization from narrow line regions of AGN. 
In doing so, we use the set of updated \citet{Feltre2016} photoionization models as presented in \citet{Mignoli2019}, developed using the version of \texttt{CLOUDY} presented in \citet{Ferland2013}. 
The original \citet{Feltre2016} model grid assumes a fixed relation between the AGN luminosity and the distance of the gas cloud from the central source, with an inner radius of $300$~pc for an AGN luminosity of $10^{45}$~erg~s$^{-1}$. 
Reducing this inner radius at a fixed AGN luminosity leads to an increase in radiation pressure.
We consider an updated version of the original \citet{Feltre2016} model grid, which includes turbulent velocity of $100$~km~s$^{-1}$ and a smaller inner radius of $90$~pc (see \citealt{Hirschmann2019,Mignoli2019,Vidal-Garcia2024}).
We summarize the main adjustable parameters below, but for full details the reader is directed to \citet{Feltre2016}.
The emitted spectrum from the AGN accretion disk is assumed to have a power law, going as $f_{\nu}\propto\nu^\alpha$ between $\lambda=0.001$ and $0.25\ \mu$m.
We will first consider $\alpha=-2.0$ (an arbitrary choice), but we will investigate the impact of allowing $\alpha$ to vary. 
We note that \citet{Feltre2016} use $\alpha=-1.2$ to $\alpha=-2.0$ and note that $\alpha=-1.7$ was inferred as the slope blueward of Ly$\alpha$ in the stacked spectrum of 53 quasars at $z\simeq 2.3$ \citep{Lusso2015}.
The ionization parameter ($U_{\rm S}$) is also an adjustable parameter in the models, defined at the Stromgren radius of the nebula. 
The hydrogen number density ($n_{\rm{H}}$) and gas metallicity ($Z$) are additional free parameters. 
The heavy element abundances ratios are the same as those adopted in \citet{Gutkin2016} and \citet{Feltre2016}, largely based on \citet{Caffau2011} for the solar abundance scale. 
All element abundances except for nitrogen are assumed to scale linearly with the oxygen abundance. 
The relative nitrogen abundance (N/O) is related to the oxygen abundance (O/H) in a manner that accounts for secondary production of nitrogen, largely following the updated prescription of \citet{Groves2004} in \citet{Gutkin2016}. 
We assume a carbon-to-oxygen (C/O) ratio equal to the solar abundance ($\log{\rm (C/O)}=-0.26$; \citealt{Asplund2009}). 
Depletion of metals onto dust grains is considered through definition of $\xi_{\rm d}$, the dust-to-heavy element mass ratio parameter, following the treatment introduced in \citet{Charlot2001}. 
In the models described below, we initially assume $\xi_{\rm d}=0.3$, but we note that assuming different $\xi_{\rm d}$ values ($0.1-0.5$) does not impact the model N~{\small V}/C~{\small IV} and N~{\small V}/He~{\small II} ratios significantly \citep[e.g.,][]{Mignoli2019}.

We show the N~{\small V}/C~{\small IV} and N~{\small V}/He~{\small II} line ratios for our fiducial set of narrow line AGN models in Figure~\ref{fig:UV_diag_alpha2p0}. 
The ionization parameter is varied between $\log{U_{\rm S}}=-5$ and $-1$, and the metallicity is varied between $Z=0.006\ Z_{\odot}$ and $4\ Z_{\odot}$. 
The N~{\small V}/C~{\small IV} and N~{\small V}/He~{\small II} ratios will be sensitive to the ionization parameter and the hardness of ionizing radiation given the relative ionization energies of each species ($77$, $54$, and $48$~eV for N~{\small V}, He~{\small II}, and C~{\small IV}). 
We show how the line ratios increase with ionization parameter in the right panel of Figure~\ref{fig:UV_diag_alpha2p0}. 
We see that N~{\small V} can be as strong as He~{\small II} and as strong as C~{\small IV} for models with ionization parameters above $\log{U_{\rm S}}=-2.5$. 
At a given ionization parameter, the highest N~{\small V}/C~{\small IV} and N~{\small V}/He~{\small II} ratios are reached for the highest metallicity models (see also \citealt{Feltre2016}). 
This is largely due to the adopted scaling between N/O and O/H. 
Even at the highest metallicities in this fiducial grid, none of the models approach the observed limits we place on the N~{\small V}/C~{\small IV} and N~{\small V}/He~{\small II} line ratios in CEERS-1025 and CEERS-7902 in Section~\ref{sec:objects}. 

Larger N~{\small V}/He~{\small II} and N~{\small V}/C~{\small IV} ratios can be obtained if the nitrogen abundance is enhanced relative to the \citet{Groves2004} relation. 
We explore the line ratios using a grid of models with two N/O values ($\log{\rm (N/O)}=0$ and $0.7$), each of which is held fixed as a function of O/H. 
At low metallicity, these N/O values represent significant nitrogen enhancements with respect to our fiducial models, as has been seen in a subset of {\it JWST}-detected galaxies. 
We consider models with the same range of metallicity, ionization parameter, and C/O as above. 
The N/O values considered here indicate relatively low C/N ratios ($\log{\rm (C/N)}=-1.1$ when $\log{\rm (N/O)}=0$ and $\log{\rm (C/N)}=-1.8$ when $\log{\rm (N/O)}=0.7$), similar to those inferred for the strong nitrogen line emitters at high redshifts \citep[e.g.,][]{Cameron2023a,Isobe2023b,Kobayashi2024,Topping2025a}. 
With the nitrogen-enhanced abundance pattern, the observed line ratios are significantly increased, particularly at low metallicities. 
We find that models with $\log{\rm (N/O)}=0.7$ are able to reproduce the observed limits at a range of ionization parameters and metallicities that are either among the lowest or among the highest in the grid. 

We note that these models also significantly boost the strength of N~{\small IV}] emission, which we do not detect. 
One potential explanation for the absence of N~{\small IV}] emission in nitrogen-enhanced clouds could be that the density is above the critical density for N~{\small IV}] ($\sim5\times10^{9}$~cm$^{-3}$ for N~{\small IV}]~$\lambda1486$ and $\sim1\times10^{6}$~cm$^{-3}$ for [N~{\small IV}]~$\lambda1483$). 
This would imply the N~{\small V}-emitting clouds are distinct from those seen in the rest of the narrow line spectrum, with densities similar to those expected from the broad line region. 
In this picture, it is conceivable that the N~{\small V} emission stems from broad line region gas that is ejected in an outflow primarily along the polar axis, perpendicular to the line of sight (see Section~\ref{sec:discussion}). 
This would enable the line emission to probe gas with small line-of-sight dispersion, producing a narrow emission line. 
If the N~{\small V}-emitting gas approaches BLR densities, then it may also contribute to the weakness of He~{\small II} emission as collisional de-excitation becomes important at such large densities \citep[e.g.,][]{Netzer1985,Mignoli2019}.

There are several other factors which may boost the N~{\small V}/He~{\small II} and N~{\small V}/C~{\small IV} ratios. 
Given that N~{\small V} requires the highest ionization energy of the three emission lines, a shallower AGN power law spectrum will boost N~{\small V} relative to C~{\small IV} and He~{\small II} (see also \citealt{Feltre2016}).
In Figure~\ref{fig:UV_diag_alpha1p2}, we present models computed assuming an AGN spectrum with a shallower power law ($\alpha=-1.2$). 
The harder spectrum will both increase the electron temperature and present more photons capable of producing NV ions. 
We show the same four panels as in Figure~\ref{fig:UV_diag_alpha2p0}, with N/O increasing with O/H (top panels) and two values of enhanced N/O (bottom panels). 
As expected, the N~{\small V}/He~{\small II} and N~{\small V}/C~{\small IV} ratios increase given the enhanced supply of $77$~eV photons. 
We find that the median N~{\small V}/He~{\small II} ratio increases by $0.15$~dex when using the shallower power law in fiducial models with fixed inner radius. 
The nitrogen-enhanced models are best able to reproduce the line ratios, with the largest ionization parameters required to produce the N~{\small V}/He~{\small II} ratios. 
These models also have low metallicities ($\lesssim0.1\ Z_{\sun}$), with high electron temperatures ($2.7-3.5\times10^4$~K) that are consistent with the measured values of CEERS-1025 (Section~\ref{sec:CEERS-1025_opt}) and CEERS-7902 (Section~\ref{sec:CEERS-7902_opt}).

The N~{\small V}/He~{\small II} and N~{\small V}/C~{\small IV} line ratios also depend on the turbulent velocity of the particles in the gas clouds responsible for the line emission. 
In the original \citet{Feltre2016} models, the turbulent velocity was set to $0$~km~s$^{-1}$. 
As the turbulent line width is increased, continuum fluorescence of the resonant lines will become more important, boosting both N~{\small V} and C~{\small IV} fluxes. 
The N~{\small V}/He~{\small II} line ratios are boosted by $0.4$~dex on average, at fixed ionization parameter and metallicity.

To summarize, the observations of CEERS-1025 and CEERS-7902 appear to indicate elevated N~{\small V}/He~{\small II} and N~{\small V}/C~{\small IV} ratios. 
We have shown that these can be reproduced with some combination of large ionization parameter and nitrogen-enhanced gas. 
This may indicate that the stellar populations formed near the nucleus have peculiar abundance patterns, perhaps from a top-heavy initial mass function \citep[e.g.,][]{Bekki2023,Kobayashi2024}. 
Alternatively, the nitrogen-enhancements could come from tidal disruption events \citep[e.g.,][]{Kochanek2016,Cameron2023a}.
The line ratios could plausibly also be boosted if the accretion disk continuum power law is shallow in the UV, or if there are significant turbulent motions in the N~{\small V}-emitting clouds. 
The detection of C~{\small III}] and not C~{\small IV} in CEERS-7902 may point to the presence of a significant contribution from a softer spectrum, potentially from massive stars in the host galaxy. 
Future investigations may also consider density-bounded models. 
If the radiation field is very hard, the N~{\small V}-emitting gas may have its carbon in a higher ionization state (i.e., C$^{4+}$), and if the clouds are density-bounded we may primarily see N~{\small V} and C~{\small V} transitions, explaining the elevated N~{\small V}/C~{\small IV} emission. 
In this picture, the lower ionization state gas may again come from different regions in the galaxy.

While the models described above can explain the constraints on N~{\small V}/He~{\small II} and N~{\small V}/C~{\small IV}, they do not provide a natural explanation for the  N~{\small V} doublet ratios in both sources.  
One effect that may alter the N~{\small V} line ratios is resonant scattering of (redshifted) Ly$\alpha$ and continuum photons by N~{\small V} ions \citep[e.g.,][]{Surdej1987,Turnshek1988,Weymann1991,Hamann1993,Krolik1998,Wang2010}. 
The separation between Ly$\alpha$ and N~{\small V} is $\sim5700$ km s$^{-1}$. 
If there is a fast outflow of dense ionized gas surrounding the broad emission line region, it is possible that both Ly$\alpha$ and continuum will be redshifted into the N~{\small V} resonance of ions in the outflow, producing N~{\small V} line emission. 
Other resonant transitions (i.e., C~{\small IV}) could also be excited by scattering in the broad absorption line region, but the presence of Ly$\alpha$ blueward of N~{\small V} will act to boost N~{\small V} relative to C~{\small IV}. 
Resonant scattering is thus an attractive explanation for observations of large N~{\small V}/C~{\small IV} and N~{\small V}/He~{\small II} ratios.

Whether resonant scattering enhances N~{\small V} in quasars remains debated. 
\citet{Hamann1996} argue that scattering is unlikely to play a significant role in boosting N~{\small V} emission since only a small fraction of Ly$\alpha$ photons would be scattered by gas in the broad absorption line region. 
However, \citet{Wang2010} have shown that N~{\small V} can be enhanced by $2-3\times$ with realistic models of the geometry and optical depth profile in the  broad absorption line (BAL) outflows. 
In their analysis, the impact of scattering on N~{\small V} emission depends on the covering factor, optical depth, and velocity profile of the outflowing gas, and the intrinsic EW of Ly$\alpha$. 
The outflow is assumed to be equatorial, providing a very large radial optical depth to Ly$\alpha$ photons. 
The scattered N~{\small V} photons are more likely to escape tangentially, which results in both a narrower line profile and viewing-angle-dependent N~{\small V} EW. 

It is also possible that such a scattering scenario may alter the N~{\small V} doublet ratio, but it requires some fine tuning. 
If the outflowing gas has maximum velocity between $5700$ and $6700$ km s$^{-1}$, redshifted Ly$\alpha$ will be more likely to excite the N~{\small V}~$\lambda1239$ transition than the N~{\small V}~$\lambda1243$ transition (here ignoring the width of Ly$\alpha$). 
This will increase the N~{\small V}~$\lambda1239$ / N~{\small V}~$\lambda1243$ doublet ratio, potentially reaching values seen in CEERS-7902.  
However, the N~{\small V} emission in CEERS-7902 does not appear significantly offset from systemic, making it difficult to understand in this framework unless the outflow is not only finely-tuned in its velocity profile but also oriented perpendicular to the line-of-sight. 
While detailed modeling may find a combination of geometries and optical depth and velocity outflow profiles which can explain the observations, such an analysis is currently beyond the scope of this paper. 
Knowledge of the kinematics of the ionized gas in the JWST-discovered AGN population is critical for interpreting resonant lines in the observed rest-frame UV spectra. 
Ultimately, a deeper spectrum capable of detecting both the N~{\small V} doublets will be required to provide a complete interpretation of the nature of N~{\small V} emission.



\begin{figure*}
\includegraphics[width=\linewidth]{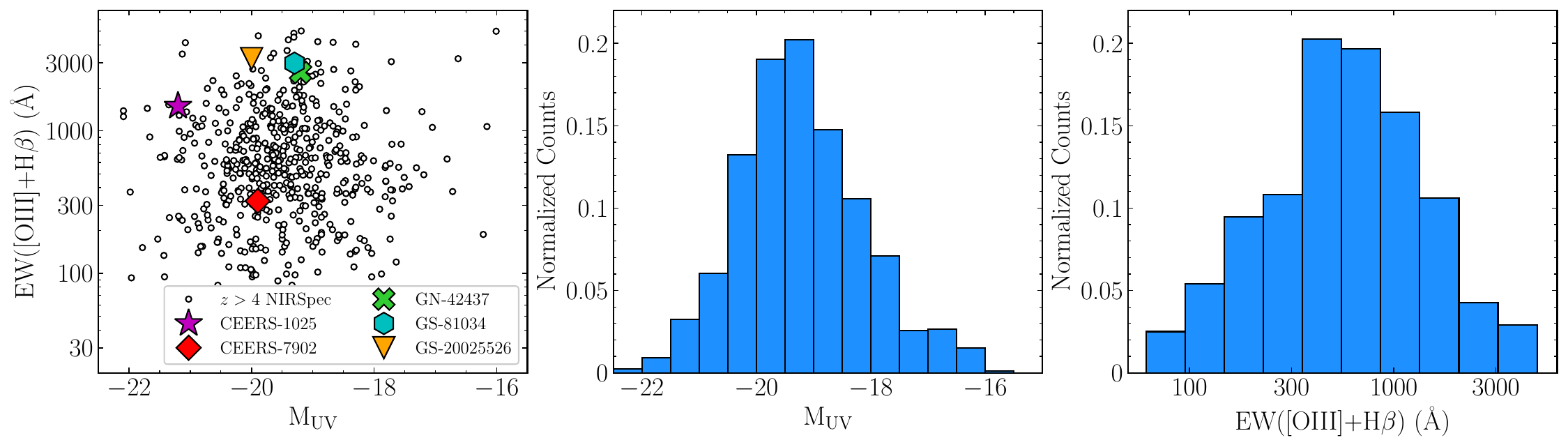}
\caption{Left panel: M$_{\rm UV}$ versus [O~{\scriptsize III}]+H$\beta$ EW for galaxies at $z>4$ with $R=1000$ or $R=2700$ NIRSpec grating spectra (black circles). We overplot objects with high ionization line detections: CEERS-1025 (N~{\scriptsize V}~$\lambda1243$, magenta star), CEERS-7902 (N~{\scriptsize V}~$\lambda1239$, red diamond), GS-20025526 (orange upside down triangle), GN-42437 (green cross; see also \citealt{Chisholm2024}), GS-81034 (cyan hexagon). Middle and right panels: M$_{\rm UV}$ and [O~{\scriptsize III}]+H$\beta$ EW distributions of $z>4$ galaxies.}
\label{fig:muv_ew}
\end{figure*}

\section{How common are very high ionization lines in {\it JWST} spectroscopy at $\lowercase{z}>4$?} \label{sec:archive}

In previous sections, we have described two galaxies with likely detections of N~{\small V} identified in the GO 4287 program. 
In this section, we aim to place galaxies with very high ionization emission lines in a broader context, characterizing how commonly these galaxies appear in $z\gtrsim 4$ samples and comparing their properties to that of the full galaxy population. 
To achieve this goal, we assemble a large database of galaxies at $z>4$ with publicly-available NIRSpec spectra and constrain the strengths of [N~{\small V}], [Ne~{\small IV}], and [Ne~{\small V}] emission lines. 
We also identify the strongest He~{\small II} emitters, noting that many of these are fully consistent with being stellar in origin. 
We choose these lines as they are most likely to be signposts of AGN photoionization (probing photons above the He$^+$ ionizing edge), although we note they are also plausibly powered by shocks. 
This complements several recent investigations of Type II AGN fractions \citep{Mazzolari2024b,Scholtz2025}, although we note our approach focuses only on the subset with high ionization emission lines.
We describe the database in Section~\ref{sec:full_sample}.
We then discuss the demographics of galaxies with high ionization lines in Section~\ref{sec:demographics}. 
We both consider the presence of these lines in the full spectroscopic sample, as well as that in the subset of our sample that are confirmed as broad line AGN with SEDs consistent with the LRD population.

\subsection{Public JWST/NIRSpec Spectra} \label{sec:full_sample}


\begin{figure*}
\includegraphics[width=\linewidth]{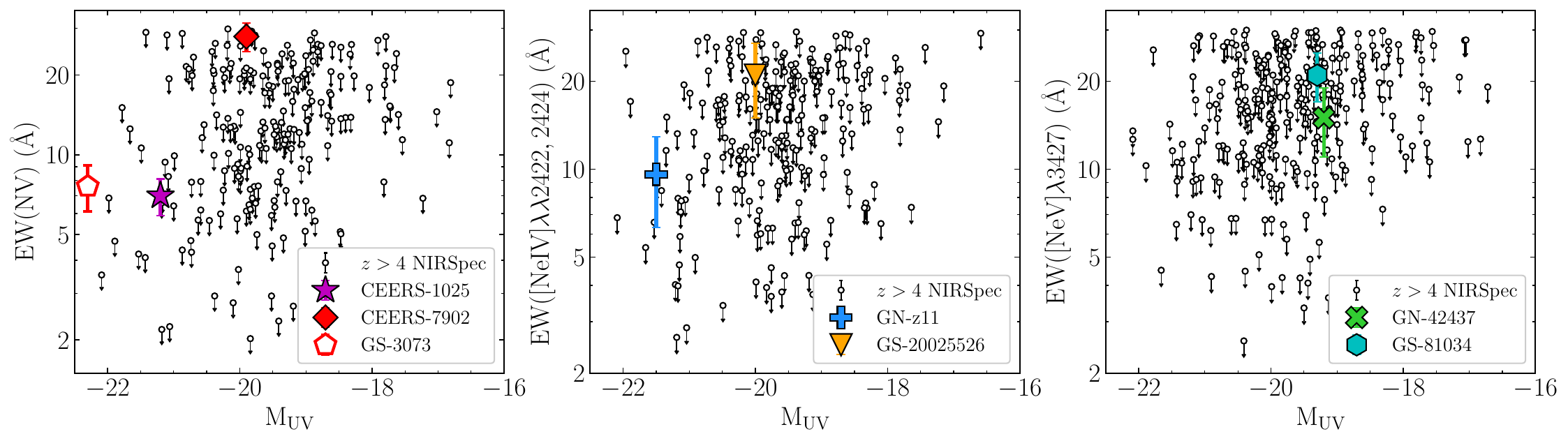}
\caption{N~{\scriptsize V}, [Ne~{\scriptsize IV}]~$\lambda\lambda2422,2424$, and [Ne~{\scriptsize V}]~$\lambda3427$ EW versus M$_{\rm UV}$ for $z>4$ galaxies with $R=1000$ or $R=2700$ NIRSpec grating spectra. We show sources with EW or $3\sigma$ EW limit $<30$~\AA\ (open black circles). For N~{\scriptsize V} (left), EW measurements of single component are presented. For [Ne~{\scriptsize IV}] (middle), we show the doublet EWs. Sources with high ionization line detections are overplotted with the same symbols in Figure~\ref{fig:muv_ew} in addition with GN-z11 (blue plux; see also \citealt{Maiolino2024a}). We also overplot the $z=5.55$ AGN GS-3073 (open red pentagon) with N~{\scriptsize V} doublet EW $=7.6\pm1.5$~\AA\ \citep{Ji2024}.}
\label{fig:ew_lim}
\end{figure*}

We construct a spectroscopic sample of $z>4$ galaxies using NIRSpec spectra obtained from the following public observations: the JWST Advanced Deep Extragalactic Survey\footnote{\url{https://jades-survey.github.io/}} (JADES, GTO 1180, 1181, PI: D. Eisenstein, GTO 1210, 1286, PI: N. L\"utzgendorf, GTO 1287, PI: K. Isaak, GO 3215, PIs: D. Eisenstein \& R. Maiolino; \citealt{Eisenstein2023a,Eisenstein2023b,Bunker2024,DEugenio2025}), the GLASS-JWST Early Release Science Program\footnote{\url{https://archive.stsci.edu/hlsp/glass-jwst}} (GLASS, ERS 1324, PI: T. Treu; \citealt{Treu2022}), CEERS, GO 1871 (PI: J. Chisholm; \citealt{Chisholm2024}), and GO 2478 (PI: D. Stark; \citealt{Topping2024,Topping2025a}) programs. 
All the NIRSpec observations were performed with the multi-object spectroscopy mode. 
The spectra were reduced following the same approaches described in \citet{Topping2025a} and Section~\ref{sec:data}. 
We refer readers to \citet{Tang2024c} and \citet{Topping2025a} for a full description of the sample selection and analysis of the spectra of the public spectroscopic sample. 

We briefly summarize the NIRSpec observations used in this section. 
Medium-resolution ($R=1000$) NIRSpec grating spectra were taken in the JADES, CEERS, and GO 4287 programs. 
The exposure time of $R=1000$ grating spectra spans from $0.9$~hour to $11.7$~hours (median $2.6$~hours) for G140M, $0.9$~hour to $6.9$~hours (median $2.6$~hours) for G235M, and $0.9$~hour to $46.7$~hours (median $2.6$~hours) for G395M. 
For a galaxy with $H=27$ at $z=6$, this median depth results in a typical $3\sigma$ rest-frame EW limit of $18$, $17$, and $28$~\AA\ for G140M, G235M, and G395M spectra, respectively. 
While this is sufficient for rest-frame optical spectroscopy, we will show in Section~\ref{sec:lrd} that this depth limits the constraints that can be placed on the weaker rest-frame UV emission lines.
High-resolution ($R=2700$) grating spectra were taken in the GLASS and GO 1871 programs, and JADES programs 1210, 1181, 1286, 1287 took G395H/F290LP spectra as well.
For the small subset of our sample with $R=2700$ spectra, the exposure time is $4.9$~hours for G140H, $4.9-14.7$~hours (median $4.9$~hours) for G235H, and $2.6-6.9$~hours (median $4.9$~hours) for G395H. 
The typical $3\sigma$ EW limit (for a $H=27$ galaxy at $z=6$) reaches to $3$, $4$, and $15$~\AA\ for G140H, G235H, and G395H spectra, respectively. 
These limits are more conducive to rest-frame UV emission line detections, but we will show in the next subsection that the number of spectra approaching these limits is still small. 

We assemble a spectroscopic database of $z\gtrsim 4$ galaxies from the aforementioned programs.
In order to identify high ionization emission lines, we focus on galaxies with precise redshift measurements using strong emission line detections. 
Most of the sources have strong rest-frame optical emission lines (e.g., H$\beta$, [O~{\small III}], H$\alpha$) in NIRSpec spectra ($z\lesssim9.5$). 
For each of these objects, we simultaneously fit the available strong optical lines with Gaussian profiles, and compute the systemic redshift using the fitted line centers. 
For a small number of galaxies at $z\gtrsim9.5$ where strong optical lines are shifted out of NIRSpec spectra, we determine the redshifts using other emission lines in the spectra (e.g., [Ne~{\small III}], [O~{\small II}], C~{\small III}]). 
We do not identify objects purely based on the Ly$\alpha$ break in this database given the need for precise redshifts. 

To ensure robust constraints on high ionization emission lines, we limit our study to a sample of galaxies with NIRSpec medium- or high-resolution ($R=1000$ or $R=2700$) grating spectra (see Section~\ref{sec:introduction}). 
From the above programs we identify $846$ such galaxies at $z>4$ with NIRSpec $R=1000$ or $R=2700$ grating spectra. 
In addition, we include the new $z>4$ galaxies identified from GO 4287 (Section~\ref{sec:data}). 
Because $5$ of the $58$ $z>4$ galaxies in GO 4287 have been observed in CEERS before, this adds $53$ new sources. 
Overall these form a sample of $899$ galaxies with NIRSpec $R=1000$ or $R=2700$ spectra at $4.0<z<14.3$. 
We measure the emission line fluxes and EWs following the same procedures described in Section~\ref{sec:data} and in \citet{Tang2024c}. 

We show the absolute UV magnitudes and [O~{\small III}]+H$\beta$ EWs of the $899$ galaxies in the spectroscopic sample in Figure~\ref{fig:muv_ew}. 
The M$_{\rm UV}$ are derived using the NIRCam photometry released by the DAWN JWST archive\footnote{\url{https://dawn-cph.github.io/dja/index.html}} team \citep{Valentino2023}. 
For the $899$ galaxies in the $z>4$ sample, the M$_{\rm UV}$ range from $-22.1$ to $-15.9$ with a median value of $-19.3$. 
The [O~{\small III}]+H$\beta$ EWs of galaxies at $z>4$ span from $62$~\AA\ to $5027$~\AA, with a median value of $691$~\AA. 
This median [O~{\small III}]+H$\beta$ EW is comparable to the average EW of photometrically-selected galaxies at high redshift at fixed M$_{\rm UV}$ \citep{Endsley2024}. 

We search for galaxies displaying N~{\small V}, [Ne~{\small IV}], He~{\small II}, or [Ne~{\small V}] emission lines in the $z>4$ spectroscopic sample by visually inspecting their 2D spectra. 
There are $851$ objects with NIRSpec $R=1000$ or $R=2700$ grating spectra covering at least one of these high ionization lines. 
We require S/N $>3$ for a line detection. 
For the non-resonant lines  ([Ne~{\small IV}], He~{\small II}, or [Ne~{\small V}]), we require candidate detections to lie 
at wavelengths within 1 instrument resolution element offset from the line center defined by the redshift of the other narrow lines ($\pm300$~km~s$^{-1}$ for $R=1000$, or $\pm111$~km~s$^{-1}$ for $R=2700$). 
For N~{\small V} emission, we allow a larger velocity offset range ($\pm1000$~km~s$^{-1}$) given its resonant nature. 
We do not consider C~{\small IV} in this paper, as a similar search  has recently been presented in \citet{Topping2025a}.

We first focus on N~{\small V}, [Ne~{\small IV}], and [Ne~{\small V}]. 
We do find a very small number of galaxies that show plausible detections. 
Aside from the two likely N~{\small V} lines presented in Section~\ref{sec:objects}, the most clear example of a robust high ionization line detection in our database is GN-42437, the [Ne~{\small V}] emitter at $z=5.587$ presented in \citet{Chisholm2024}. 
In our reduction, the [Ne~{\small V}] line is detected at S/N=4 with a modest EW (EW $=15\pm4$~\AA), similar to what was reported in \citet{Chisholm2024}. 
We also tentatively detect the [Ne~{\small IV}]~$\lambda2422,2424$ emission in GN-z11 ($z=10.604$, EW $=9.6\pm3.3$~\AA) that was first reported in \citet{Maiolino2024a}. 
The S/N of the line is low ($3$), but we consider this a plausible high ionization emission feature. 
The origin of the line emission in GN-z11 remains debated (e.g., \citealt{Bunker2023,Maiolino2024a,Alvarez-Marquez2025}).
Whether the low S/N [Ne~{\small IV}] detection reflects AGN activity or is consistent with a combination of shocks and massive stars is not yet clear. 
In addition to these sources, we find two new lower significance (S/N $=3-3.5$) potential detections of [Ne~{\small IV}] or [Ne~{\small V}] emission. 
We consider these as very tentative, and note that deeper data will ultimately be required to verify whether or not the emission lines are real. 
One of the putative sources is GS-81034 ($z=5.390$), where we find a potential emission feature (S/N $=3$) near 
[Ne~{\small V}]~$\lambda3427$ (EW $=15\pm5$~\AA). 
The other source is GS-20025526 ($z=7.951$) where we note the presence of a low significance (S/N $=3.5$) emission feature (EW $=21\pm6$~\AA) near the expected location of [Ne~{\small IV}]~$\lambda\lambda2422,2424$. 
We verify that each of these two tentative detections is also detected in two independent reductions that are publicly available (DJA; \citealt{Heintz2024,deGraaff2025b}, and JADES; \citealt{Bunker2024,DEugenio2025}).
We summarize our measurements of the four sources with plausible high ionization line emission in  Appendix~\ref{sec:new_high_ion}. 
Together with the two N~{\small V} emitting galaxies identified in GO 4287 (Section~\ref{sec:objects}), these sources form a small subset of galaxies with either confirmed or potential high ionization lines in archival NIRSpec medium- or high-resolution spectra at $z>4$. 

There are six sources at $z>4$ (GS-9422, GS-58975, GS-202208, RXCJ2248-ID, A1703-zd2, and A1703-zd6) with detectable He~{\small II} emission, all previously reported in the literature \citep{Cameron2024,Topping2024,Topping2025a,Curti2025,Scholtz2025}. 
We measure He~{\small II} EW $=2-10$~\AA\ in our reductions of these spectra, consistent with what has been presented in  previous investigations of the sources. 
Each of the grating spectra also show O~{\small III}]~$\lambda\lambda1661,1666$ emission lines, allowing calculation of the O~{\small III}]/He~{\small II} flux ratios. 
The measured ratios ($=1.0-4.5$) are consistent with expectations for stellar photoionization  \citep{Feltre2016}. 
While we cannot rule out AGN activity, it seems likely that there is  significant stellar contribution to the ionizing spectrum. 
In the following subsection, we will primarily focus on what the detections of N~{\small V}, [Ne~{\small IV}], and [Ne~{\small V}] ($816$ objects with NIRSpec spectra covering at least one of these lines) suggest for the prevalence of very hard radiation fields at $z\gtrsim 4$.

The vast majority ($810$) galaxies do not show the high ionization emission lines described above in their NIRSpec spectra. 
But not all non-detections are equally constraining given the widely varying continuum strengths and sensitivity limits.
We place $3\sigma$ upper limits on the EW of high ionization lines in the $810$ galaxies without detections. 
There are $383$, $476$, and $632$ galaxies with NIRSpec spectra covering N~{\small V}, [Ne~{\small IV}], and [Ne~{\small V}], respectively. 
We derive the 25th-50th-75th percentiles of $3\sigma$ upper limits of N~{\small V} (each single component of the doublet) EW $=11$, $21$, and $48$~\AA. 
For [Ne~{\small IV}] ([Ne~{\small V}]) line, the 25th-50th-75th percentiles of $3\sigma$ EW upper limit are $12$, $21$, and $43$~\AA\ ($17$, $32$, and $60$~\AA), respectively. 
It is clear that many of the existing spectra are only able to detect very strong high ionization line emission, but there is a subset which allow useful constraints on the incidence of hard radiation fields. 

\subsection{Fraction of Galaxies with High Ionization Lines} \label{sec:demographics}

In this subsection, we aim to quantify the fraction of $z>4$ galaxies showing very high ionization emission lines in existing spectra. 
By focusing on lines above the He$^+$-ionizing edge, we aim to constrain the incidence of non-stellar photoionization in existing spectroscopic samples. 
As motivated in the previous subsection, we will focus on N~{\small V}, [Ne~{\small IV}], and [Ne~{\small V}] detections.
We first discuss the general population at $z>4$ in Section~\ref{sec:general_pop} and then consider the subset of LRDs with confirmed broad lines in Section~\ref{sec:lrd}. 

\subsubsection{The General Population at $z\gtrsim 4$} \label{sec:general_pop}

As described in Section~\ref{sec:full_sample}, many of the $816$ galaxies in our $z>4$ database do have deep enough spectra to place a sensitive limit on the high ionization line EW ($\simeq10$~\AA, Section~\ref{sec:objects}; see also, e.g., \citealt{Chisholm2024}). 
In order to put robust constraints on the fraction of high ionization line emitters, we primarily focus on galaxies with $3\sigma$ EW limit $<10$~\AA\ for at least one of the high ionization lines. 
This limits us to a sample of $185$ galaxies at $z>4$. 
Among this subset, there are $87$, $97$, and $57$ objects with deep enough spectra to reach N~{\scriptsize V}, [Ne~{\scriptsize IV}], or [Ne~{\scriptsize V}] EW limits of $<10$~\AA, respectively. 

Using this database, we can compute the detection rate of each individual high ionization line.
For the $87$ sources with $3\sigma$ N~{\small V} EW upper limits below $10$~\AA, two present N~{\small V} emission (CEERS-1025 and CEERS-7902). 
This indicates that the fraction of galaxies with strong N~{\small V} emission is fairly small ($2/87=2.3^{+3.0}_{-1.5}\%$; uncertainties estimated using the statistics for small numbers of events; \citealt{Gehrels1986}). 
The percentage of galaxies presenting [Ne~{\small IV}] ([Ne~{\small V}]) emission line is $1/97=1.0^{+2.4}_{-0.9}\%$ ($1/57=1.8^{+4.1}_{-1.5}\%$). This increases to $2/97=2.1^{+2.7}_{-1.3}\%$ ($2/57=3.5^{+4.5}_{-2.3}\%$) for both lines if we include the tentative detections described above. 
If we consider a slightly larger EW threshold ($<15$~\AA\ at $3\sigma$), the fraction of high ionization line emitters decreases to $4/308=1.3^{+1.0}_{-0.6}\%$ ($6/308=1.9^{+1.2}_{-0.8}\%$ if including tentative detections). 
These results suggest that strong very high ionization lines appear rarely in $z\gtrsim4$ galaxies, with only a few percent of the population exhibiting N~{\small V}, He~{\small II}, [Ne~{\small IV}], or [Ne~{\small V}] with EW $>10$~\AA.

The fraction of high ionization line emitters is comparable if we limit our sample to galaxies in our spectroscopic database without known broad emission lines (i.e., removing broad line AGN). 
To quantify the fraction in this case, we remove the $19$ broad line AGN (with Balmer emission line FWHM $>1000$~km~s$^{-1}$) in our $z>4$ sample. 
This leaves $166$ galaxies with $3\sigma$ EW limit $<10$~\AA\ for at least one of the N~{\small V}, [Ne~{\small IV}], or [Ne~{\small V}] line. 
We find $5$ plausible high ionization lines in this sample, resulting a fraction of $3.0^{+2.0}_{-1.3}\%$. 
This fraction is somewhat lower than that reported in several recent studies \citep{Mazzolari2024b,Scholtz2025}, but we note those investigations also included galaxies identified by rest-frame optical diagnostics. 
A deeper spectroscopic database will ultimately be required to better establish the fraction of high ionization lines and the viability of low S/N detections. 
But even with the existing database, it appears clear that very strong (EW $>10$~\AA) high ionization lines probing above 54 eV are fairly rare in the overall galaxy population at $z\gtrsim 4$. 
We note that our high ionization line fraction is roughly comparable to that measured in ground-based spectra at $z\simeq 2-3$ \citep{Hainline2011}, however direct comparison to this value is difficult given the different magnitudes of the respective samples (the \citealt{Hainline2011} sample is significantly brighter) and the different selection criteria that have been used for identifying high ionization detections. 
As a result, the relative AGN fractions may also reflect mass-dependent differences between the two samples.

\subsubsection{High Ionization Lines in LRDs} \label{sec:lrd}


\begin{figure*}
\includegraphics[width=\linewidth]{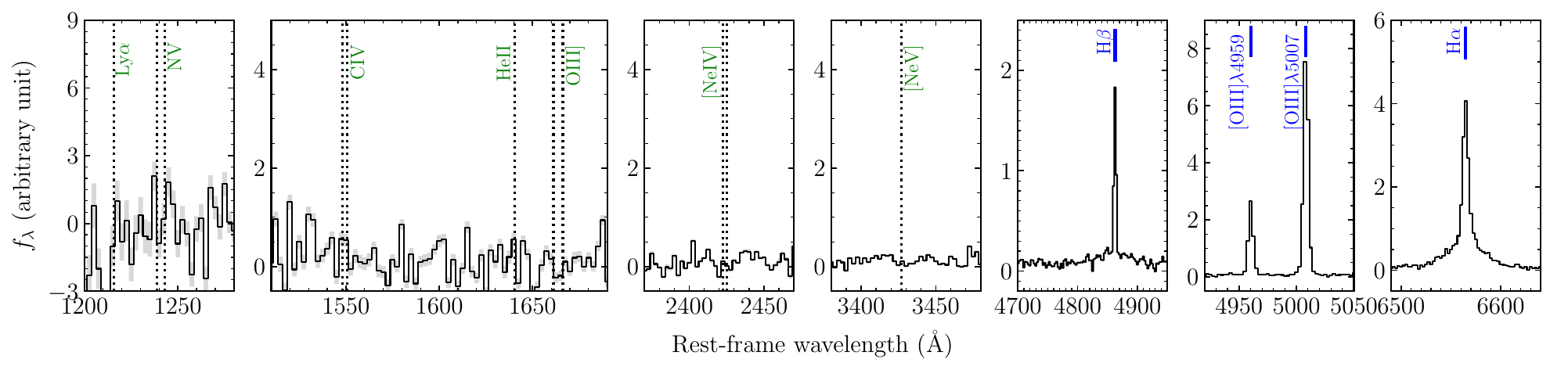}
\caption{Composite NIRSpec grating spectrum of the $18$ LRDs at $z>4$. High ionization emission lines (N~{\scriptsize V}, C~{\scriptsize IV}, He~{\scriptsize II}, [Ne~{\scriptsize IV}], [Ne~{\scriptsize V}]) are not detected, which are marked by green text. We show detections of narrow forbidden [O~{\scriptsize III}]~$\lambda4959$ and $\lambda5007$ lines, as well as H$\beta$ and H$\alpha$ emission line with narrow and broad components.}
\label{fig:stack_lrd}
\end{figure*}

To place the detection of N~{\small V} in CEERS-7902 in context, we investigate how frequently strong high ionization emission in BL AGN in the $z>4$ spectroscopic sample introduced in Section~\ref{sec:full_sample}. 
We specifically focus on the subset of those that would be classified as LRDs, but we note that this represents a subset of the total BL AGN population \citep{Hainline2025}. 

To assemble the sample of LRDs with deep grating rest-frame UV spectra, we apply the  selection criteria similar to that  applied in \citet{Matthee2024}, requiring detection of broad H$\beta$ or H$\alpha$ emission (S/N $>5$) with FWHM $>1000$~km~s$^{-1}$. 
Then we select red and compact sources from the BL AGN following the color criteria in \citet{Kokorev2024} and \citet{Labbe2025}. 
For galaxies at $z<6$, we apply color cuts of F115W $-$ F150W $<0.8$, F200W $-$ F277W $>0.7$, and F200W $-$ F356W $>1.0$. 
For galaxies at higher redshift ($z>6$), we apply F150W $-$ F200W $<0.8$, F277W $-$ F356W $>0.6$, and F277W $-$ F444W $>0.7$. 
To ensure we select compact sources, we further require that the F444W images are spatially unresolved using the ratio of F444W flux densities measured between $0''.4$ and $0''.2$ apertures $f_{\rm F444W}(0''.4)/f_{\rm F444W}(0''.2)<1.7$. 
With these selection criteria, we identify $18$ LRD BL AGN among the $899$ galaxies in our grating rest-frame UV database.
We list the $18$ LRDs at $z>4$ with NIRSpec grating spectra in Table~\ref{tab:lrd}, whose NIRSpec spectra have also been presented in literature \citep{Harikane2023,Kocevski2023,Maiolino2024b,Wang2024,Juodzbalis2025,Kocevski2025,Taylor2025}. 

\begin{deluxetable*}{ccccccccc}
\tablecaption{LRDs confirmed as BL AGN with NIRSpec medium- and high-resolution grating sampling rest-frame UV.}
\tablehead{
ID & PID & R.A. & Decl. & $z_{\rm spec}$ & M$_{\rm UV}$ & Line & FWHM$_{\rm broad}$ & References \\
 & & (deg) & (deg) & & (AB mag) & & (km~s$^{-1}$) &
}
\startdata
CEERS-672 & 1345 & $214.889677$ & $+52.832977$ & $5.6669$ & $-17.7$ & H$\alpha$ & $1394\pm333$ & (1,8,9) \\
CEERS-746 & 1345 & $214.809142$ & $+52.868484$ & $5.6227$ & $-18.3$ & H$\alpha$ & $1630\pm177$ & (1,2,3,8,9) \\
CEERS-2782 & 1345 & $214.823453$ & $+52.830281$ & $5.2410$ & $-20.0$ & H$\alpha$ & $2084\pm305$ & (1,2,9) \\
CEERS-7902 & 4287 & $214.983038$ & $+52.956205$ & $6.9827$ & $-19.9$ & H$\beta$ & $3394\pm259$ & (4,6,9,this work) \\
CEERS-10444 & 4287 & $214.892231$ & $+52.877651$ & $6.6835$ & $-19.9$ & H$\beta$ & $3816\pm128$ & (3,6,8,9,this work) \\
GN-954 & 1181 & $189.151972$ & $+62.259639$ & $6.7597$ & $-19.9$ & H$\alpha$ & $1839\pm84$ & (4,7) \\
GN-1093 & 1181 & $189.179742$ & $+62.224629$ & $5.5946$ & $-18.0$ & H$\alpha$ & $2046\pm216$ & (4,7) \\
GN-11836 & 1181 & $189.220587$ & $+62.263675$ & $4.4087$ & $-19.3$ & H$\alpha$ & $1472\pm182$ & (4,7) \\
GN-20621 & 1181 & $189.122515$ & $+62.292850$ & $4.6814$ & $-17.7$ & H$\alpha$ & $1900\pm253$ & (4,7) \\
GN-53757 & 1181 & $189.269778$ & $+62.194208$ & $4.4475$ & $-19.1$ & H$\alpha$ & $2044\pm106$ & (4,7) \\
GN-61888 & 1181 & $189.168016$ & $+62.217013$ & $5.8738$ & $-18.6$ & H$\alpha$ & $1528\pm131$ & (4,7) \\
GN-73488 & 1181 & $189.197396$ & $+62.177233$ & $4.1327$ & $-18.9$ & H$\alpha$ & $2225\pm51$ & (4,7) \\
GS-9597 & 1180 & $53.166115$ & $-27.772043$ & $6.3057$ & $-19.0$ & H$\alpha$ & $2565\pm228$ & (3) \\
GS-9515 & 1210 & $53.132763$ & $-27.801838$ & $4.6487$ & $-18.6$ & H$\alpha$ & $1598\pm166$ & (4,7) \\
GS-13704 & 1210 & $53.126677$ & $-27.818102$ & $5.9198$ & $-19.3$ & H$\alpha$ & $2288\pm278$ & (3,4,7,8) \\
GS-38562 & 1286 & $53.135876$ & $-27.871755$ & $4.8222$ & $-18.9$ & H$\alpha$ & $1909\pm52$ & (7) \\
GS-172975 & 1286 & $53.087715$ & $-27.871211$ & $4.7421$ & $-17.5$ & H$\alpha$ & $1538\pm140$ & (3,7,8) \\
GS-204851 & 1286 & $53.138541$ & $-27.790279$ & $5.4825$ & $-19.1$ & H$\alpha$ & $2010\pm69$ & (5,7) \\
\enddata
\tablecomments{Column ``PID'' shows the JWST program ID. Column ``Line'' lists the Balmer emission line with the highest S/N broad component identified in NIRSpec spectra, whose broad line widths are listed in column ``FWHM$_{\rm broad}$''. \\
{\bf References}: (1) \citet{Harikane2023}; (2) \citet{Kocevski2023}; (3) \citet{Kokorev2024}; (4) \citet{Maiolino2024b}; (5) \citet{Matthee2024}; (6) \citet{Wang2024}; (7) \citet{Juodzbalis2025}; (8) \citet{Kocevski2025}; (9) \citet{Taylor2025}.}
\label{tab:lrd}
\end{deluxetable*}

We characterize the strength of narrow rest-frame UV high ionization emission lines (N~{\small V}, He~{\small II}, [Ne~{\small IV}], and [Ne~{\small V}]) in the NIRSpec spectra of these $18$ LRDs. 
The NIRSpec $R=1000$ or $R=2700$ grating spectra cover N~{\small V} emission lines in $8$ LRDs. 
Four of these eight LRDs have robust N~{\small V} constraints with $3\sigma$ EW upper limit $<10$~\AA, and we detect N~{\small V} in one of them (CEERS-7902, see Section~\ref{sec:CEERS-7902}). 
There are $15$ LRDs with NIRSpec grating spectra covering He~{\small II}. 
No He~{\small II} emission is seen in the seven LRDs with deep enough spectra to detect EW $=10$~\AA\ line emission. 
There are $13$ and $16$ LRDs with NIRSpec grating spectra covering [Ne~{\small IV}] and [Ne~{\small V}], respectively. Neither line is seen in this sample.  
The [Ne~{\small IV}] and [Ne~{\small V}] EW limits are relatively large, with only $3$ LRDs with [Ne~{\small IV}] EW limit $<10$~\AA\ and $4$ LRD with [Ne~{\small V}] EW limit $<10$~\AA. 
These results underscore how few grating spectra have yet reached depths necessary to detect high ionization lines at EW $=10$~\AA. 
Larger samples of deep $R=1000$ spectra will soon allow improved constraints, but based on the presence of N~{\small V} emission in the small existing sample, it is possible that high ionization lines may be relatively common in LRD spectra.

We also search for narrow C~{\small IV}~$\lambda\lambda1548,1551$ emission.
Fifteen of the galaxies have NIRSpec $R=1000$ or $R=2700$ grating spectra with wavelength ranges covering C~{\small IV}.  
We detect unresolved C~{\small IV} doublet emission in one LRD, GN-954 ($z=6.7597$). 
This line was previously reported in the census of C~{\small IV} emission presented in \citet{Topping2025a}. 
We measure a C~{\small IV} doublet EW $=31.5\pm5.6$~\AA, consistent within $1\sigma$ of the EW reported in \citet{Topping2025a} (EW $=23.3^{+3.8}_{-3.7}$~\AA). 
This on its own is not enough to verify the line is powered by AGN photoionization, but it is plausible that a non-stellar hard radiation field contributes given the broad lines detected in the rest-frame optical. 
For the remaining $14$ LRDs with C~{\small IV} constraints, we do not identify C~{\small IV} emission, with $7$ of them having $3\sigma$ EW upper limits $<10$~\AA. 
While the statistics are poor, this suggests a C~{\small IV} detection rate of $1/8=12.5^{+23.7}_{-10.4}\%$. 

Above we have focused on narrow rest-frame UV emission lines in LRDs. 
We can also constrain the presence of broad permitted UV lines in the spectra of the $18$ LRDs in grating rest-frame UV coverage. 
\citet{Lambrides2024} have recently shown that no broad high ionization UV lines are present in a sample of $8$ LRDs. 
We find the same result is present with our sample of $18$ LRDs. 
Using the spectra, we can constrain the upper limits of luminosities of broad high ionization emission lines (N~{\small V}, C~{\small IV}, He~{\small II}, [Ne~{\small IV}], [Ne~{\small V}]). 
To do so, we assume a broad line width that is the same as that measured from broad Balmer emission in the rest-frame optical (Table~\ref{tab:lrd}). 
Then for each high ionization line, we compute the $1\sigma$ upper limit by integrating the error spectrum in quadrature using a spectral window spanning $2\times$ FWHM$_{\rm broad}$.  
For the $18$ LRDs, the median $3\sigma$ upper limits of the broad emission line luminosities are $2.6\times10^{41}$~erg~s$^{-1}$ for N~{\small V}, C~{\small IV}, or He~{\small II}, and $1.5\times10^{41}$~erg~s$^{-1}$ for [Ne~{\small IV}] and $1.1\times10^{41}$~erg~s$^{-1}$ for [Ne~{\small V}].

To place more stringent constraints on the broad high ionization emission lines, we create a composite spectrum of $18$ LRDs.
We first shift individual LRD spectra to the rest-frame using the systemic redshifts measured from narrow [O~{\small III}] lines. 
Each spectrum is then interpolated to a common rest-frame wavelength scale of $2.5$~\AA\ ($R\simeq500-1000$ at rest-frame UV) and normalized by its H$\beta$ luminosity. 
Finally the spectra are median stacked and the composite spectrum is multiplied by the median H$\beta$ luminosity of the $18$ LRDs. 
The composite spectrum also does not show broad UV lines.
To constrain the broad emission luminosities, we again assume a line width equal to the average broad Balmer emission line width (FWHM $=1950$~km~s$^{-1}$) and integrate the composite error spectrum in quadrature. 
The $3\sigma$ upper limits of broad line luminosities estimated from the composite spectrum are $4.4\times10^{40}$~erg~s$^{-1}$ for C~{\small IV} or He~{\small II}, and $1.5\times10^{40}$~erg~s$^{-1}$ for [Ne~{\small V}]. 
The upper limits estimated from our composite LRD spectrum are consistent with the upper limits for systems presented in \citet{Lambrides2024}. 

Detailed modeling of the high ionization  broad line luminosities is beyond the scope of this paper. 
However, \citet{Lambrides2024} have presented expected high ionization broad line luminosities expected for slim disk models with weak EUV radiation fields resulting from the effects of photon trapping. 
Based on their results, we may have expected to detect broad emission lines in the UV at the composite luminosity limit, even after accounting for potential dust extinction effects.
This result may suggest that the intrinsic UV spectrum of LRDs is (on average) weak, as suggested in \citet{Lambrides2024}.  However, it appears plausible that some LRDs may indeed power hard radiation fields given the presence of narrow high ionization lines. 
And others may have broad UV lines strongly attenuated by H~{\small I} scattering if there is complete line-of-sight coverage by dense hydrogen that is excited to the $n=2$ level  (e.g., \citealt{Inayoshi2025,Ji2025,Naidu2025}). 
Such dense hydrogen would effectively scatter away UV photons from the disk and the broad line region.
Deeper rest-frame UV spectra of individual LRDs should reveal whether a subset may show broad UV lines. 
\citet{Ubler2023} present a spectrum of a $z=5.55$ AGN with broad He~{\small II}~$\lambda1640$ emission exceeding the luminosity limit of our composite. 
As larger rest-frame UV spectroscopic databases emerge, it should be possible to determine whether the properties of broad line AGN with high ionization lines differ from those without.


\section{Discussion} \label{sec:discussion}

We have quantified the presence of broad and narrow high ionization lines in a large sample $z\gtrsim 4$ {\it JWST} spectra. 
The visibility of high ionization UV lines will depend both on the intrinsic EUV radiation field and on the opacity provided by gas and dust.
Our work supports earlier indications that high ionization UV lines are rare \citep{Lambrides2024}, both in LRDs and in the general population. 
However, we do detect narrow N~{\small V} emission in two $z>6$ galaxies, one of which is an LRD previously shown to have broad H$\beta$ emission \citep{Wang2024,Kocevski2025}, while the other shows only narrow lines in the rest-frame UV and optical. 
Neither shows broad UV emission lines. 
These narrow line detections add to a small but growing database of galaxies with high ionization lines \citep{Labbe2025,Topping2025a,Treiber2025}. 
This may indicate that at least a subset of the population does power intense EUV ionizing spectra. 
With these sources, we can begin to explore the conditions which facilitate the high ionization features. 


\begin{figure}
\centering
\includegraphics[width=\linewidth]{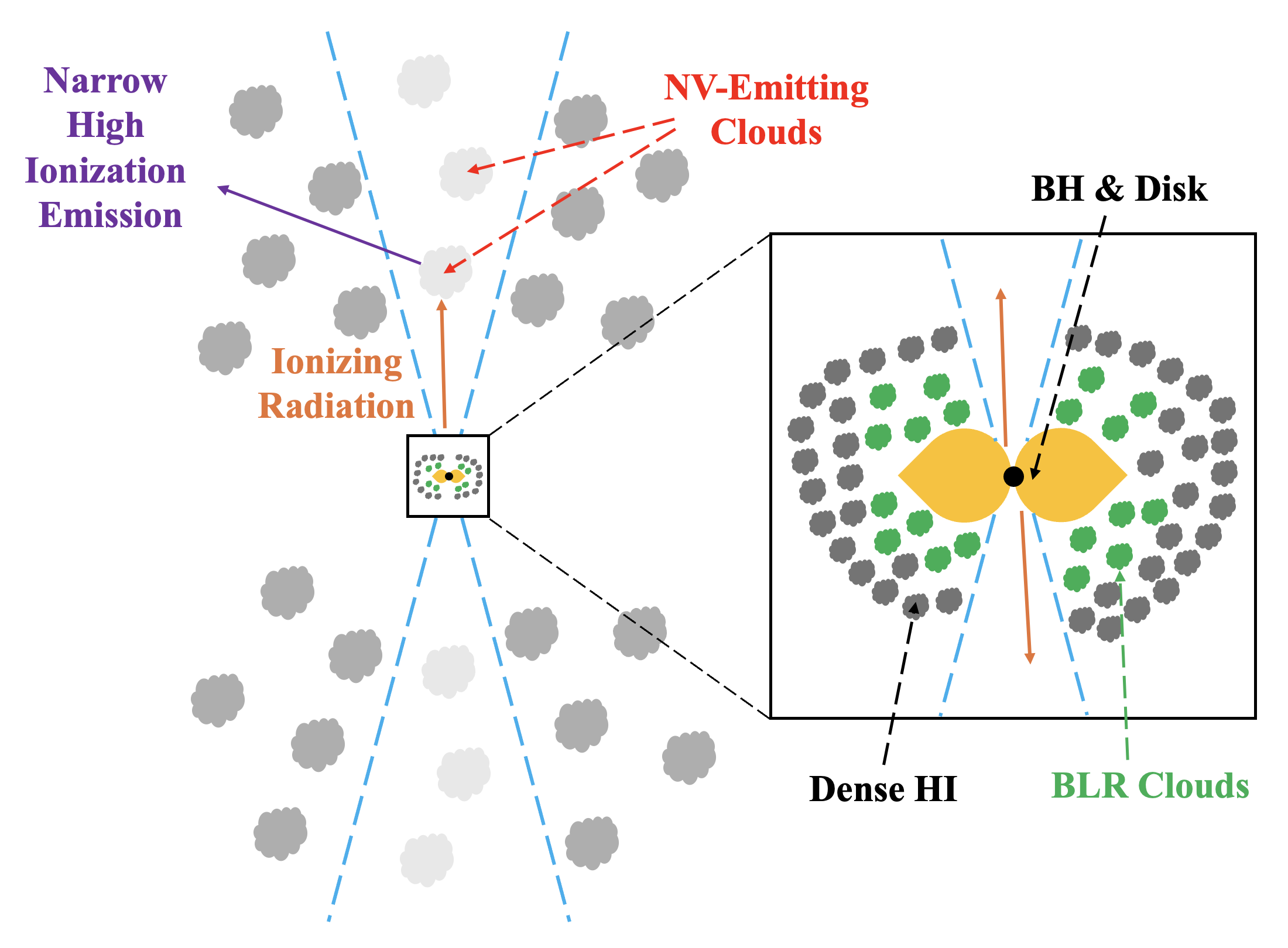}
\caption{Schematic illustrating a scenario for AGN photoionization of narrow line clouds in LRDs with significant line-of-sight covering of extremely dense neutral hydrogen (dark grey) in the vicinity of  the BLR clouds (green) and the disk (orange), resulting in Balmer series absorption in the rest-frame optical spectrum. If the line-of-sight covering fraction of dense gas and BLR clouds is large, the narrow line region may still be ionized along the polar axis, perpendicular to the line-of-sight. The narrow high ionization lines may be further amplified if the line-of-sight covering fraction of dense hydrogen and the BLR is reduced.}
\label{fig:lrd_schematic}
\end{figure}

In the case of the N~{\small V}-emitting LRD presented in Section~\ref{sec:CEERS-7902} (CEERS-7902), we observe a Balmer break and H$\beta$ absorption. 
In the context of recent papers \citep{deGraaff2025a,Inayoshi2025,Ji2025,Naidu2025}, this indicates the presence of extremely dense neutral gas around the accretion disk in the N~{\small V} emitter, with neutral hydrogen populating the $n=2$ level. 
\citet{Naidu2025} suggest that the dense gas likely uniformly covers the disk and broad line region in the case of the LRD MoM-BH$^{\star}$-1. 
In that system, uniform neutral gas coverage is critical for powering the strong Balmer break ($f_{\nu,4050}/f_{\nu,3670}=7.7^{+2.3}_{-1.4}$) and producing a symmetric H$\beta$ absorption feature (c.f. \citealt{Juodzbalis2025}). 
A similar picture may also apply to The Cliff, another LRD with a very strong ($f_{\nu,4050}/f_{\nu,3670}=6.9^{+2.8}_{-1.5}$) Balmer break \citep{deGraaff2025a}. 

The presence of narrow N~{\small V} emission and Balmer series absorption in the spectrum of CEERS-7902 may indicate that its dense neutral gas (with hydrogen in its $n=2$ level) does not uniformly cover its nucleus, allowing ionizing radiation  to escape to the narrow line region (see also \citealt{Lin2025}). 
One possibility is that the ionizing radiation is collimated into a funnel-like geometry, with the opening angle directed along the rotation axis, as predicted for AGNs undergoing super-Eddington accretion \citep[e.g.,][]{Sikora1981,Madau1988,Madau2024,Pacucci2024}. 
It has recently been noted that the supercritical disk geometry and anisotropic radiation field may help explain the absence of X-rays in most LRDs \citep{Pacucci2024,Madau2025}. 
This configuration may also be useful for understanding the narrow and broad line spectra \citep[e.g.,][]{Wang2014}.
If our viewing angle  is oriented far from the rotation axis (i.e., closer to edge-on), the ionization cone would be oriented closer to perpendicular to the line-of-sight.
The AGN continuum would ionize gas clouds within the ionization cone at larger impact parameters. 
At these distances, the line-of-sight opacity to UV photons is much lower.  
As a result, we should be able to observe narrow high ionization UV lines even if our line-of-sight to the central disk is significantly obscured by dense neutral gas, as in the case in CEERS-7902 (Figure~\ref{fig:lrd_schematic}). 
In this configuration, the broad UV lines would be significantly attenuated by absorption from hydrogen atoms in the $n=2$ level, potentially explaining how we might see narrow high ionization line emission without seeing broad UV lines.  
Of course, the prominence of the narrow high ionization lines would be additionally boosted if the line-of-sight covering fraction of dense neutral gas is not uniform, allowing a larger volume of the narrow line region to be irradiated with the AGN spectrum. 
In practice, our ability to detect the narrow high ionization lines will further depend on the contribution from massive stars in the host galaxy, as significant underlying UV continuum emission would reduce the emission line EWs.

It is possible that the N~{\small V}-emitting clouds are distinct from those producing the rest of the narrow line spectrum. This may help explain why CEERS-7902 presents N~{\small V} and C~{\small III}] without any hint of C~{\small IV} emission. 
In this picture the C~{\small III}] emission would come from a different region than the N~{\small V} emission, possibly from clouds outside of the AGN ionization cone where the softer stellar ionizing spectrum dominates. 
We find hints of this picture from the AGN photoionization models introduced in Section~\ref{sec:line_ratios}. 
Models reproducing the observed N~{\small V}/C~{\small IV} and N~{\small V}/He~{\small II} limits predict larger N~{\small V}/C~{\small III}] ratios ($\simeq6$) than that measured in CEERS-7902 ($3.4$), likely suggesting different origins of the N~{\small V} and C~{\small III}] emission.
In Section~\ref{sec:line_ratios}, we suggested that the observed N~{\small V}/C~{\small IV} and N~{\small V}/He~{\small II} ratios could be explained if the clouds were nitrogen-enhanced with densities similar to that of the broad line region, as this would help explain the absence of N~{\small IV}] emission. 
One possibility is that the N~{\small V} emission comes from broad line region gas that has been ejected in an outflow along the polar axis perpendicular to the line-of-sight, such that the majority of their motion is not in the line-of-sight. 
The high ionization clouds may be more centrally concentrated than those dominating narrow line emission from lower ionization species.


\begin{figure*}
\includegraphics[width=\linewidth]{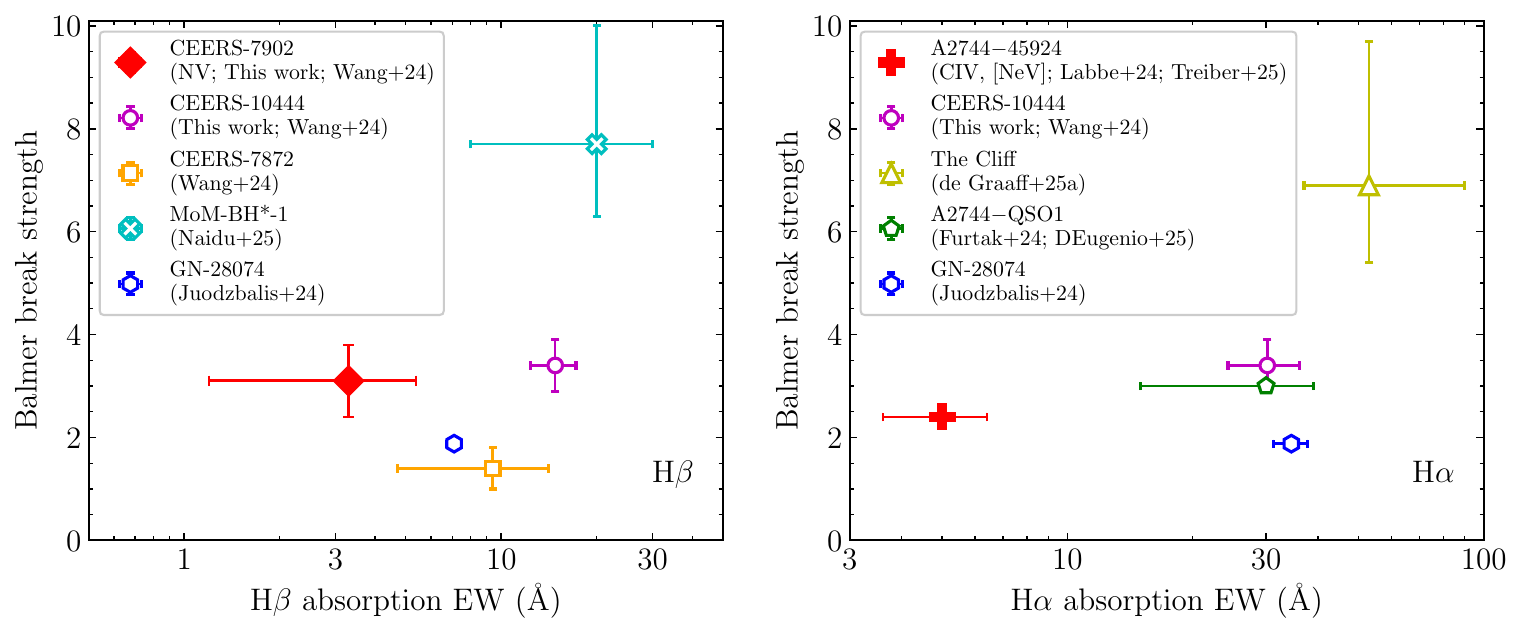}
\caption{Balmer break strengths $f_{\nu,4050}/f_{\nu,3670}$ versus Balmer absorption line EWs of LRDs presented in this work and literature \citep{Furtak2024,Juodzbalis2024,Labbe2024,Wang2024,deGraaff2025a,DEugenio2025,Naidu2025,Treiber2025}. The left panel shows LRDs with H$\beta$ absorption lines, and the right panel shows those with H$\alpha$ absorption lines. The N~{\scriptsize V} emitter CEERS-7902 is marked by solid red diamond, and the LRD with prism C~{\scriptsize IV} and [Ne~{\scriptsize V}] detections identified in \citet{Labbe2024} and \citet{Treiber2025}is marked by solid red cross. Other LRDs without high ionization emission line detections are shown as open symbols.}
\label{fig:lrd_balmer}
\end{figure*}

We may also expect to see additional signatures of AGN photoionization in the narrow line spectra of LRDs with high ionization lines. 
As described in Section~\ref{sec:CEERS-7902_opt}, the rest-frame optical spectrum of CEERS-7902 shows [O~{\small III}]$\lambda4363$/H$\gamma$ vs. [Ne~{\small III}]/[O~{\small II}] ratios that are not expected in star formation galaxies \citep{Mazzolari2024a}, likely reflecting the enhanced heating provided by the AGN continuum on the narrow-line clouds. 
We also report a tentative detection of the low ionization line [O~{\small I}]~$\lambda6302$ in CEERS-7902 (and a clear detection in the LRD CEERS-10444), often seen in partially-ionized regions of AGNs. 
The implied [O~{\small I}]~$\lambda6302$/H$\alpha$ ratio of CEERS-7902 is consistent with expectations for AGN 
photoionization \citep[e.g.,][]{Kewley2001,Feltre2016,Mazzolari2024b}. 
Given that we see both of these features alongside a strong Balmer break and H$\beta$ absorption, we may again expect that a hard ionizing radiation field from the AGN is escaping to the narrow line region in spite of the indications of dense neutral gas absorption along the line-of-sight.

In the above discussion, we have assumed the high ionization lines are powered by AGN photoionization. 
However, the highly-ionized species may also originate from other processes. 
Fast radiative shocks could create the conditions necessary to power the high ionization lines without requiring a hard EUV AGN spectrum \citep[e.g.,][]{Allen2008,Izotov2012,Alarie2019}. 
If turbulent velocities are as large as suggested in some recent investigations \citep[e.g.,][]{Naidu2025}, such shocks may not surprising. 
However, the UV line emission produced by the shocks would potentially be absorbed by the dense neutral gas implied by the Balmer break (with hydrogen predominantly in the $n=2$ level) if the line-of-sight covering fraction is near-unity, unless there is significant line production in the outer optically thin regime.
It is also possible that massive stars in the host galaxies contribute. 
While it is unlikely that normal stellar populations supply the photon budget responsible for producing species with ionization energies well above the He$^+$ ionizing edge (i.e., N~{\small V}, [Ne~{\small V}]), it is conceivable that the dense conditions in LRDs produce a population of ionizing sources unlike that seen at lower redshifts. 
Deeper spectra should enable these alternative possibilities to be tested in greater detail.

As larger samples of deep rest-frame UV spectra are obtained, it will be possible to compare the properties of LRDs exhibiting high ionization UV lines with those that do not. 
We may expect high ionization narrow lines to be less likely in LRDs with the disk and broad line region blanketed by dense H~{\small I} gas \citep{Inayoshi2025}. 
Perhaps surprisingly in that context, the existing database reveals that two of the LRDs with high ionization narrow line detections (CEERS-7902 and A2744-45924) do have Balmer breaks and Balmer series absorption features. 
However, as shown in Figure~\ref{fig:lrd_balmer}, these two LRDs have weaker H$\beta$ or H$\alpha$ absorption lines and  lower amplitude Balmer Breaks than those recently presented in \citet{Naidu2025} and \citet{deGraaff2025a}.
This may indicate that galaxies with high ionization lines have a lower line-of-sight covering fraction of extremely dense neutral gas. 
It may also point to a different intrinsic spectrum or a larger opening angle for the funnel geometry, both of which may relate to the accretion physics \citep[e.g.,][]{Wang2014,Jiang2019}.
More detailed photoionization modeling jointly fitting the high ionization lines and the Balmer series absorption 
is beyond the scope of this paper, but should yield valuable insights as higher quality spectra are obtained in the future. 
If the geometry described above is valid, it should also be possible to identify AGNs viewed pole-on, along the funnel-like ionization cone. 
These sources should be much brighter in the UV, resulting in lower high ionization EWs \citep{Madau2025}. 
If there are gas clouds experiencing fast outflows along the polar axis, we should similarly see evidence in the small subset of AGNs viewed closer to pole-on.

Of course, the broad line LRDs described above are only a subset of the $z\gtrsim 4$ AGN population \citep[e.g.,][]{Hainline2025}. 
Identification and characterization of low luminosity AGNs that do not show broad lines will help place the LRD and broad line AGN population in context. 
It is well known that traditional rest-frame optical narrow line diagnostics struggle in identifying Type II AGN at very high redshift owing to the low metallicity and extreme ionization gas conditions that become common in star forming galaxies \citep[e.g.,][]{Masters2014,Coil2015,Ubler2023,Mazzolari2024b,Scholtz2025}. 
New diagnostics focused on high ionization or temperature-sensitive emission lines will help \citep[e.g.,][]{Feltre2016,Mazzolari2024a}, but if the UV radiation field of AGNs is intrinsically weak or ionizing photons are strongly attenuated prior to reaching the narrow line region, these diagnostics will also be incomplete. 

Progress requires continued efforts to identify Type II AGN as {\it JWST} spectroscopic samples grow in number.
We have presented an N~{\small V} emitter at $z=8.7$ without broad lines in its rest-frame optical spectrum. 
This galaxy has a large [O~{\small III}]+H$\beta$ EW ($1682$~\AA) and a very blue continuum slope ($\beta=-2.5$), comparable to other systems dominated by very young stellar populations at similar redshifts. 
This galaxy is one of a small sample of Type II AGN candidates at $z\gtrsim 4$ discovered with {\it JWST} spectroscopy \citep{Ubler2023,Mazzolari2024b,Chisholm2024,Silcock2024,Scholtz2025}. 
Our investigation of $851$ public NIRSpec grating spectra suggests that the fraction of $z\gtrsim 4$ galaxies with high ionization lines (N~{\small V}, He~{\small II}, [Ne~{\small IV}], [Ne~{\small V}]) is $2.2^{+1.7}_{-1.0}\%$, where here this census refers only to lines with EW $>10$~\AA. 
This percentage is broadly similar to that measured in ground-based follow-up of UV-selected galaxies at $z\simeq2-3$ \citep[e.g.,][]{Hainline2011}. 
We emphasize that this number should be considered as merely the fraction of galaxies with very high ionization lines and not as an AGN fraction. 
Deeper spectra will be required to validate low S/N detections and isolate the most likely powering mechanism (i.e., shocks, AGN). 

Our results suggest that a subset of $z>4$ AGN may have elevated nitrogen emission line strengths, with detectable N~{\small V} in cases where C~{\small IV} and He~{\small II} are not present. 
While the precise physics driving the strength of the N~{\small V} emission is not clear, it is possible that a nitrogen enhanced abundance pattern may be partially responsible (see Section~\ref{sec:line_ratios}), as has been seen in other AGN with {\it JWST} \citep[e.g.,][]{Ji2024,Napolitano2024,Tripodi2024,Isobe2025}. 
This may suggest that the star formation conditions near the nucleus produce unusual abundance patterns, perhaps reflecting a very top heavy initial mass function capable of producing nitrogen-enhancements \citep[e.g.,][]{Bekki2023,Kobayashi2024}. 
Alternatively we may be seeing the effects of gas produced via tidal disruption events \citep[e.g.,][]{Kochanek2016,Cameron2023a}. 
One of the key questions is whether this is a common phase in the JWST-detected AGN population. 
The frequency of the high ionization lines will depend on the radiation field, the gas covering fraction and kinematics, and the strength of the underlying UV continuum.
To-date only four LRDs have been observed with NIRSpec grating spectroscopy capable of detecting N~{\small V} at EW as low as $10$~\AA. 
For reference, the composite of Type II AGN in \citet{Hainline2011} has N~{\small V} EW $=5.6$~\AA. 
Deeper spectra sensitive to narrow N~{\small V} emission with EW down to $5$~\AA\ (in both the general population and in the LRDs) may uncover a larger sample of AGNs with hard radiation fields and (potentially) nitrogen-enhanced gas conditions.


\section{Summary} \label{sec:summary}

We present a search for very high ionization emission lines (N~{\small V}, [Ne~{\small IV}], [Ne~{\small V}]) in $851$ galaxies at $z>4$ with {\it JWST}/NIRSpec medium- or high-resolution ($R=1000$ or $R=2700$) grating spectra. 
The dataset includes new $R=2700$ G140H/F100LP spectra of $58$ sources in the EGS field observed through the GO 4287 program. 
From the new observations we identify two likely narrow N~{\small V} emission line detections, providing indications of hard radiation fields usually associated with AGNs.
We summarize our findings below. 

1. We detect narrow N~{\small V}~$\lambda1243$ emission  (EW $=7.0\pm1.1$~\AA) in the new G140H spectrum of CEERS-1025 ($z=8.7166$), a galaxy previously-confirmed via rest-frame optical emission lines in the CEERS program \citep{Nakajima2023,Tang2023}. 
The rest-frame optical spectrum reveals no broad lines, and the SED is similar to that seen in star forming galaxies dominated by young stellar populations. 
No C~{\small IV} or He~{\small II} emission is seen in the rest-frame UV spectrum of CEERS-1025.

2. We report rest-frame UV spectroscopy of two LRDs, both previously confirmed to show broad hydrogen emission lines and Balmer series absorption, as expected for broad line AGN surrounded by extremely dense neutral gas with hydrogen populating its $n=2$ level. 
The rest-frame UV spectrum of one of the two LRDs (CEERS-7902, $z=6.9827$) shows strong, narrow N~{\small V}~$\lambda1239$   (EW $=27.9\pm3.4$~\AA) and C~{\small III}], but C~{\small IV} and He~{\small II} are not detected. 
The other LRD (CEERS-10444 at $z=6.6836$) also shows C~{\small III}], but N~{\small V} is not covered in the G140H spectrum, and C~{\small IV} and He~{\small II} are not detected. 
In spite of the indications of dense neutral gas absorption around the central disk and broad line region, both LRDs show the potential influence of AGN photoionization on the narrow emission lines (i.e., [O~{\small I}]~$\lambda6302$, [O~{\small III}]~$\lambda4363$), suggesting the ionizing continuum is able to escape along channels to impact gas in the narrow line region. 

3. The detections of N~{\small V} without C~{\small IV} or He~{\small II} suggest elevated N~{\small V}/C~{\small IV} ($>1.4$) and N~{\small V}/He~{\small II} ($>2.6$) ratios, as have been seen in broad lines of nitrogen-loud quasars.
We show that the line ratios can be reproduced by a combination of nitrogen-enhanced and turbulent gas, with a large ionization parameter and a shallow power law spectrum. 
The nitrogen enhancement may indicate the presence of top-heavy initial mass functions or tidal disruption events in the vicinity of the nucleus. 
The N~{\small V} flux ratios may be further boosted by resonant scattering of Ly$\alpha$ photons which have been redshifted into the N~{\small V} resonance by a fast outflow, as has been previously suggested for strong nitrogen emitting quasars \citep[e.g.,][]{Hamann1996,Wang2010}. 
We suggest that if the densities of the N~{\small V}-emitting gas are similar to that of the broad line region, it may help explain the weak He~{\small II} emission and N~{\small IV}] non-detection. 
The doublet ratios in both N~{\small V} emitters are inconsistent with the expected intrinsic ratios, as may be expected if the line emission is scattered through outflowing or inflowing gas. 
The resonant scattering of Ly$\alpha$ photons may further alter N~{\small V} line ratios, while density-bounded gas may additionally play a role in driving the line ratios. 
Deeper spectra capable of detecting both components of the N~{\small V} doublets will ultimately be required to fully understand the nature of N~{\small V} emission of CEERS-1025 and CEERS-7902.

4. We explore the incidence of very high ionization emission (N~{\small V}, He~{\small II}, [Ne~{\small IV}], [Ne~{\small V}]) in a sample of $851$ continuum-selected $z>4$ galaxies and $18$ $z>4$ LRDs with moderate or high resolution NIRSpec observations covering the rest-frame UV. 
We find that $2.2^{+1.7}_{-1.0}\%$ of the $z>4$ galaxy population has spectra with plausible detections of the very high ionization lines with EW $>10$~\AA. 
This fraction is comparable to that seen at lower redshifts in continuum-selected samples \citep{Hainline2011}, although a direct comparison is challenging given existing current EW limits. 
Most LRDs do not yet have spectra deep enough to constrain the full suite of narrow high ionization UV lines (N~{\small V}, C~{\small IV}, He~{\small II}, [Ne~{\small IV}], [Ne~{\small V}]) to EWs of $10$~\AA, but the very limited samples suggest these lines are present in a subset of LRDs ($25.0^{+37.1}_{-20.8}\%$ for N~{\small V}, $12.5^{+23.7}_{-10.4}\%$ for C~{\small IV}). 
However, these results  suggest that high ionization lines are weak in the majority of LRDs, potentially signally a softer ionizing spectrum or blanketing of the radiation field by dense neutral gas and dust. 

5. The presence of narrow high ionization lines in LRDs with Balmer series absorption suggests that the extremely dense neutral hydrogen gas may not uniformly blanket the disk in all cases, enabling sightlines through which 
hard photons can be transmitted to the narrow line region. 
It is conceivable that the line-of-sight covering fraction of the dense neutral gas is non-unity. 
Alternatively, the ionizing radiation may largely escape through an excavated funnel along the polar axis, oriented perpendicular to the line-of-sight. 
The dense gas along the line-of-sight may significantly attenuate UV emission lines from the broad line region (if hydrogen is excited to the $n=2$ level), potentially contributing to the weakness of the broad high ionization lines. 
Larger spectroscopic samples should reveal whether LRDs with high ionization lines differ in their properties from those that do not have high ionization features, allowing this picture to be tested in greater detail.


\section*{Acknowledgment}

The authors acknowledge the anonymous referee for insightful comments which improved the manuscript. 
The authors would like to thank Linhua Jiang, Jianwei Lyu, Marta Volonteri, Feige Wang, and Huiyuan Wang for useful discussions.
We also thank St\'{e}phane Charlot and Jacopo Chevallard for providing access to the \texttt{BEAGLE} tool used for SED fitting analysis. 
MT acknowledges funding from the \textit{JWST} Arizona/Steward Postdoc in Early galaxies and Reionization (JASPER) Scholar contract at the University of Arizona. 
DPS acknowledges support from the National Science Foundation through the grant AST-2109066. 
CAM acknowledges support by the European Union ERC grant RISES (101163035), Carlsberg Foundation (CF22-1322), and VILLUM FONDEN (37459). The Cosmic Dawn Center (DAWN) is funded by the Danish National Research Foundation under grant DNRF140. 
LW acknowledges support from the JWST/NIRCam contract to the University of Arizona, NAS5-02015, and support from the National Science Foundation Graduate Research Fellowship under Grant No. DGE-2137419.

This work is based on observations made with the NASA/ESA/CSA \textit{James Webb Space Telescope}. 
The data were obtained from the Mikulski Archive for Space Telescopes at the Space Telescope Science Institute, which is operated by the Association of Universities for Research in Astronomy, Inc., under NASA contract NAS 5-03127 for \textit{JWST}. 
These observations are associated with program GO 4287, and the following public-available programs GTO 1180, 1181, 1210, 1286, 1287, and GO 3215 (JADES, doi: 10.17909/8tdj-8n28; \citealt{Rieke2023_doi}), ERS 1324 (GLASS, doi: 10.17909/kw3c-n857; \citealt{Treu2023_doi}), ERS 1345 and DDT 2750 (CEERS, doi: 10.17909/z7p0-8481; \citealt{Finkelstein2023_doi}), GO 1871, as well as GO 4233 (RUBIES). 
The authors acknowledge the JADES, GLASS, CEERS, and RUBIES teams led by Daniel Eisenstein \& Nora L\"uetzgendorf, K. Isaak, Tommaso Treu, Steven L. Finkelstein, Pablo Arrabal Haro, and Anna de Graaff \& Gabriel Brammer for developing their observing programs. 
This research is also based in part on observations made with the NASA/ESA \textit{Hubble Space Telescope} obtained from the Space Telescope Science Institute, which is operated by the Association of Universities for Research in Astronomy, Inc., under NASA contract NAS 5–26555. 
Part of the data products presented herein were retrieved from the Dawn \textit{JWST} Archive (DJA). 
DJA is an initiative of the Cosmic Dawn Center, which is funded by the Danish National Research Foundation under grant DNRF140. 
This work is based in part upon High Performance Computing (HPC) resources supported by the University of Arizona TRIF, UITS, and Research, Innovation, and Impact (RII) and maintained by the UArizona Research Technologies department. 

%

\vspace{5mm}





\appendix

\section{Galaxies with High Ionization Emission Lines at $z>4$} \label{sec:new_high_ion}


\begin{figure*}
\includegraphics[width=\linewidth]{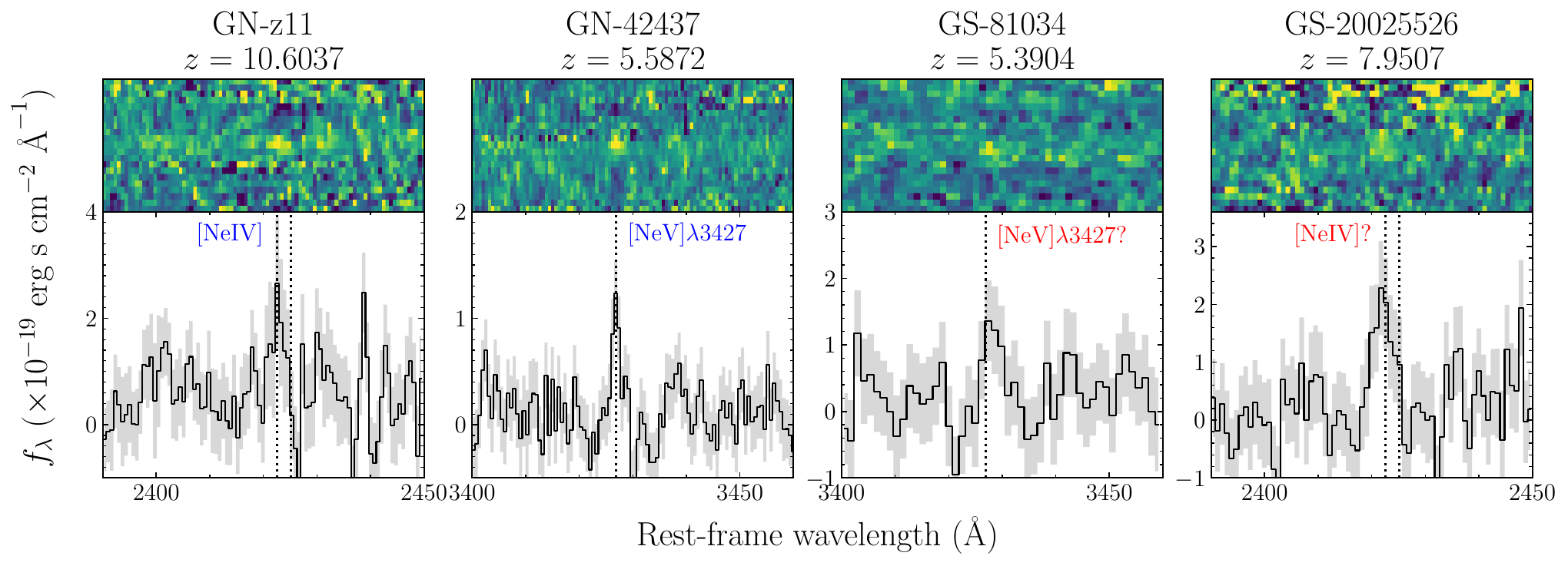}
\caption{Plausible detections (S/N $\simeq3-4$) of high ionization emission lines ([Ne~{\scriptsize V}] and [Ne~{\scriptsize IV}]) in the NIRSpec spectra of GN-z11, GN-42437, GS-81034, and GS-20025526.}
\label{fig:high_ionization_spec}
\end{figure*}

\subsection{GN-z11}

GN-z11 is a bright (M$_{\rm UV}=-21.5$) galaxy at $z=10.604$, with NIRSpec spectra (obtained from JADES GTO 1181) have been well studied previously \citep{Bunker2023}. 
We derive a systemic redshift $z_{\rm sys}=10.6037$ using the [Ne~{\small III}]~$\lambda3869$ and H$\gamma$ lines presented in the G395M spectrum, consistent with the value reported in \citet{Bunker2023}. 
Using this redshift, we detect unresolved [Ne~{\small IV}]~$\lambda\lambda2422,2424$ (S/N $=3$) in the G235M spectrum. 
The [Ne~{\small IV}] detection in GN-z11 was initially presented in \citet{Maiolino2024a}, here we present our measurements for completeness. 
We measure a line flux $=3.8\pm1.3\times10^{-19}$~erg~s$^{-1}$~cm$^{-2}$, consistent with the value ($3.14\pm0.65\times10^{-19}$~erg~s$^{-1}$~cm$^{-2}$) reported in \citet{Maiolino2024a}. 
We derive a [Ne~{\small IV}] EW $=9.6\pm3.3$~\AA. 

\subsection{GN-42437}

GN-42437 is at $z_{\rm sys}=5.5872$ with M$_{\rm UV}=-19.2$, with high-resolution ($R=2700$) NIRSpec grating spectra obtained from program 1871 \citep{Chisholm2024}. 
We determine the systemic redshift from the [O~{\small III}] doublet and H$\alpha$ emission lines in the G395H spectrum, which is consistent with the redshift measured in \citet{Chisholm2024}. 
Based on this redshift we find an emission line (S/N $=4$) in the G235H spectrum with peak at $22573.2$~\AA, consistent with the line center of [Ne~{\small V}]~$\lambda3427$ ($-6\pm52$~km~s$^{-1}$). 
The [Ne~{\small V}] detection of GN-42437 was initially reported in \citet{Chisholm2024}. 
We measure a [Ne~{\small V}] line flux $1.92\pm0.45\times10^{-19}$~erg~s$^{-1}$~cm$^{-2}$ and EW $=15\pm4$~\AA, consistent with those reported in \citet{Chisholm2024} (line flux $=2.35\pm0.34\times10^{-19}$~erg~s$^{-1}$~cm$^{-2}$, EW $=11\pm2$~\AA). 

\subsection{GS-81034}

GS-81034 (R.A. $=53.088307$, Decl. $=-27.840416$) is at $z_{\rm sys}=5.3904$ with M$_{\rm UV}=-19.3$. 
Its medium-resolution ($R=1000$) NIRSpec grating spectra were obtained from JADES program 1286. 
The systemic redshift of GS-81034 is determined from the H$\beta$, [O~{\small III}], and H$\alpha$ emission lines presented in the G395M spectrum. 
Using this redshift we identify a tentative (S/N $=3$) emission feature in the G235M spectrum, with peak ($21902.4$~\AA) consistent with the line center of [Ne~{\small V}]~$\lambda3427$ ($+42\pm107$~km~s$^{-1}$). 
We measure a [Ne~{\small V}] EW $=15\pm5$~\AA\ for GS-81034. 
We also examine whether the NIRSpec spectra reduced by other teams present this detection. 
We find that both the GS-81034 spectra reduced by the DJA and the JADES team show similar emission features at the line center of [Ne~{\small V}]~$\lambda3427$.
We search for other high ionization emission lines in the spectra of GS-81034. 
We do not detect N~{\small V} or C~{\small IV}, He~{\small II} emission, placing a $3\sigma$ upper limit of EW $<5$~\AA\ for each of these lines.

\subsection{GS-20025526}

GS-20025526 (R.A. $=53.099439$, Decl. $=-27.880489$) is a bright galaxy (M$_{\rm UV}=-20.0$) at $z_{\rm sys}=7.9507$, with $R=1000$ NIRSpec spectra obtained from JADES GTO 1286. 
The systemic redshift is determined from the H$\beta$ and [O~{\small III}] emission lines in the G395M spectrum. 
Based on this redshift we detect an $3.5\sigma$ emission feature in the G235M spectrum with a peak at $21676.2$~\AA. 
This is consistent with the expected position of the unresolved [Ne~{\small IV}]~$\lambda\lambda2422,2424$ doublet.
We measure the [Ne~{\small IV}] doublet EW $=21\pm6$~\AA. 
The similar [Ne~{\small IV}] emission feature is also detected in the spectra reduced by the JADES team. 
In the spectra reduced by DJA, the emission feature is still presence but with a lower S/N ($\simeq1.5$).]
In the G140M spectrum of GS-20025526, we detect Ly$\alpha$ emission with EW $=34\pm5$~\AA, with peak velocity offset $=+426\pm96$~km~s$^{-1}$. 
We do not detect N~{\small V} emission, placing $3\sigma$ upper limit EW $<6$~\AA\ for each component of the doublet.


\bibliography{NIRSpec_high_ionization_lines}{}
\bibliographystyle{aasjournal}



\end{document}